\documentclass[twocolumn]{aastex63}
\usepackage{amsmath}
\usepackage{graphicx}
\usepackage{float}
\usepackage{verbatim}
\usepackage{subfigure}

\graphicspath{{figures/}}

\newcommand{\spt}{SPT}
\newcommand{\sptsz}{SPT-SZ}
\newcommand{\sptpol}{SPTpol}
\newcommand{\sptthree}{SPT-3G}
\newcommand{\act}{ACT}
\newcommand{\actpol}{ACTPol}
\newcommand{\planck}{\textit{Planck}}
\newcommand{\xxl}{XXL}
\newcommand{\xmm}{\textit{XMM-Newton}}
\newcommand{\erosita}{eROSITA}
\newcommand{\lcdm}{\ensuremath{\Lambda \rm{CDM}}}
\newcommand{\mfive}{\ensuremath{M_{500c}}}
\newcommand{\msun}{\ensuremath{M_\odot h^{-1}_{70}}}
\newcommand{\msunnoh}{\ensuremath{M_\odot}}
\newcommand{\uk}{\ensuremath{\mu \rm{K} ^2}}
\newcommand{\Tcmb}{\ensuremath{T_{\textsc{cmb}}}}
\newcommand{\fsz}{\ensuremath{f_{\textsc{sz}}}}
\newcommand{\ysz}{\ensuremath{y_{\textsc{sz}}}}
\newcommand{\Asz}{\ensuremath{A_{\textsc{sz}}}}
\newcommand{\um}{\ensuremath{\mu}m}
\newcommand{\spitzer}{{\sl Spitzer}}

\newcommand{\ncandidates}{89}
\newcommand{\nconfirmed}{81}
\newcommand{\npersqdeg}{0.86}
\newcommand{\nwithz}{79}
\newcommand{\nwithphotoz}{66}
\newcommand{\nspecz}{13}
\newcommand{\nzgtone}{15}
\newcommand{\nzgtpeight}{23}
\newcommand{\pctgtone}{18\%}
\newcommand{\pctgtpeight}{29\%}
\newcommand{\massmedian}{\ensuremath{2.7 \times 10^{14} \msun}}
\newcommand{\massmin}{\ensuremath{1.8 \times 10^{14} \msun}}
\newcommand{\massmax}{\ensuremath{8.3 \times 10^{14} \msun}}
\newcommand{\maxz}{1.38}
\newcommand{\nwithfollowup}{84}
\newcommand{\nconfirmednoz}{2}
\newcommand{\nfalsedetincat}{7}
\newcommand{\nbrightstar}{1}
\newcommand{\catpurity}{92\%}
\newcommand{\sptszmatch}{26}

\newcommand{\nfirstrep}{29}     
\newcommand{\npersqdegsptsz}{0.22}
\newcommand{\minsig}{4.6} 
\newcommand{\mcomplete}{\ensuremath{2.6 \times 10^{14} \msun}}

\newcommand{\nfdet}{8} 
\newcommand{\fdrpercent}{91\%} 

\newcommand{\spitzerpct}{91\%}  

\newcommand{\webaddress}{\url{http://pole.uchicago.edu/public/data/sptsz-clusters}}

\begin{document}

\title{Galaxy Clusters Selected via the Sunyaev-Zel'dovich Effect in the \sptpol\ 100-Square-Degree Survey}

\shortauthors{N.~Huang, L.~E.~Bleem, B.~Stalder, et al.}
\author{N.~Huang} \affiliation{Department of Physics, University of California, Berkeley, CA, USA 94720}
\author{L.~E.~Bleem} \affiliation{High Energy Physics Division, Argonne National Laboratory, 9700 S. Cass Avenue, Argonne, IL, USA 60439} \affiliation{Kavli Institute for Cosmological Physics, University of Chicago, 5640 South Ellis Avenue, Chicago, IL, USA 60637}
\author{B.~Stalder} \affiliation{LSST, 950 N Cherry Ave, Tucson, AZ} \affiliation{Harvard-Smithsonian Center for Astrophysics, 60 Garden Street, Cambridge, MA, USA 02138}
\author{P.~A.~R.~Ade} \affiliation{Cardiff University, Cardiff CF10 3XQ, United Kingdom}
\author{S.~W.~Allen} \affiliation{Kavli Institute for Particle Astrophysics and Cosmology, Stanford University, 452 Lomita Mall, Stanford, CA 94305, USA} \affiliation{Department of Physics, Stanford University, 382 Via Pueblo Mall, Stanford, CA 94305, USA} \affiliation{SLAC National Accelerator Laboratory, 2575 Sand Hill Road, Menlo Park, CA 94025}
\author{A.~J.~Anderson} \affiliation{Fermi National Accelerator Laboratory, MS209, P.O. Box 500, Batavia, IL 60510}
\author{J.~E.~Austermann} \affiliation{NIST Quantum Devices Group, 325 Broadway Mailcode 817.03, Boulder, CO, USA 80305}
\author{J.~S.~Avva} \affiliation{Department of Physics, University of California, Berkeley, CA, USA 94720}
\author{J.~A.~Beall} \affiliation{NIST Quantum Devices Group, 325 Broadway Mailcode 817.03, Boulder, CO, USA 80305}
\author{A.~N.~Bender} \affiliation{High Energy Physics Division, Argonne National Laboratory, 9700 S. Cass Avenue, Argonne, IL, USA 60439} \affiliation{Kavli Institute for Cosmological Physics, University of Chicago, 5640 South Ellis Avenue, Chicago, IL, USA 60637}
\author[0000-0002-5108-6823]{B.~A.~Benson} \affiliation{Fermi National Accelerator Laboratory, MS209, P.O. Box 500, Batavia, IL 60510} \affiliation{Kavli Institute for Cosmological Physics, University of Chicago, 5640 South Ellis Avenue, Chicago, IL, USA 60637} \affiliation{Department of Astronomy and Astrophysics, University of Chicago, 5640 South Ellis Avenue, Chicago, IL, USA 60637}
\author{F.~Bianchini} \affiliation{School of Physics, University of Melbourne, Parkville, VIC 3010, Australia}
\author{S.~Bocquet} \affiliation{Department of Physics, Ludwig-Maximilians-Universit\"{a}t,Scheinerstr.\ 1, 81679 M\"{u}nchen, Germany} \affiliation{High Energy Physics Division, Argonne National Laboratory, 9700 S. Cass Avenue, Argonne, IL, USA 60439} \affiliation{Kavli Institute for Cosmological Physics, University of Chicago, 5640 South Ellis Avenue, Chicago, IL, USA 60637}
\author{M.~Brodwin} \affiliation{Department of Physics and Astronomy, University of Missouri, 5110 Rockhill Road, Kansas City, MO 64110, USA}
\author{J.~E.~Carlstrom} \affiliation{Kavli Institute for Cosmological Physics, University of Chicago, 5640 South Ellis Avenue, Chicago, IL, USA 60637} \affiliation{Department of Physics, University of Chicago, 5640 South Ellis Avenue, Chicago, IL, USA 60637} \affiliation{High Energy Physics Division, Argonne National Laboratory, 9700 S. Cass Avenue, Argonne, IL, USA 60439} \affiliation{Department of Astronomy and Astrophysics, University of Chicago, 5640 South Ellis Avenue, Chicago, IL, USA 60637} \affiliation{Enrico Fermi Institute, University of Chicago, 5640 South Ellis Avenue, Chicago, IL, USA 60637}
\author{C.~L.~Chang} \affiliation{Kavli Institute for Cosmological Physics, University of Chicago, 5640 South Ellis Avenue, Chicago, IL, USA 60637} \affiliation{High Energy Physics Division, Argonne National Laboratory, 9700 S. Cass Avenue, Argonne, IL, USA 60439} \affiliation{Department of Astronomy and Astrophysics, University of Chicago, 5640 South Ellis Avenue, Chicago, IL, USA 60637}
\author{H.~C.~Chiang} \affiliation{Department of Physics, McGill University, 3600 Rue University, Montreal, Quebec H3A 2T8, Canada} \affiliation{School of Mathematics, Statistics \& Computer Science, University of KwaZulu-Natal, Durban, South Africa}
\author{R.~Citron} \affiliation{University of Chicago, 5640 South Ellis Avenue, Chicago, IL, USA 60637}
\author{C.~Corbett~Moran} \affiliation{University of Chicago, 5640 South Ellis Avenue, Chicago, IL, USA 60637} \affiliation{TAPIR, Walter Burke Institute for Theoretical Physics, California Institute of Technology, 1200 E California Blvd, Pasadena, CA, USA 91125}
\author{T.~M.~Crawford} \affiliation{Kavli Institute for Cosmological Physics, University of Chicago, 5640 South Ellis Avenue, Chicago, IL, USA 60637} \affiliation{Department of Astronomy and Astrophysics, University of Chicago, 5640 South Ellis Avenue, Chicago, IL, USA 60637}
\author{A.~T.~Crites} \affiliation{Kavli Institute for Cosmological Physics, University of Chicago, 5640 South Ellis Avenue, Chicago, IL, USA 60637} \affiliation{Department of Astronomy and Astrophysics, University of Chicago, 5640 South Ellis Avenue, Chicago, IL, USA 60637} \affiliation{California Institute of Technology, MS 249-17, 1216 E. California Blvd., Pasadena, CA, USA 91125}
\author{T.~de~Haan} \affiliation{Department of Physics, University of California, Berkeley, CA, USA 94720} \affiliation{Physics Division, Lawrence Berkeley National Laboratory, Berkeley, CA, USA 94720}
\author{M.~A.~Dobbs} \affiliation{Department of Physics, McGill University, 3600 Rue University, Montreal, Quebec H3A 2T8, Canada} \affiliation{Canadian Institute for Advanced Research, CIFAR Program in Cosmology and Gravity, Toronto, ON, M5G 1Z8, Canada}
\author{W.~Everett} \affiliation{Department of Astrophysical and Planetary Sciences, University of Colorado, Boulder, CO, USA 80309}
\author{B.~Floyd} \affiliation{Department of Physics and Astronomy, University of Missouri, 5110 Rockhill Road, Kansas City, MO 64110, USA}
\author{J.~Gallicchio} \affiliation{Kavli Institute for Cosmological Physics, University of Chicago, 5640 South Ellis Avenue, Chicago, IL, USA 60637} \affiliation{Harvey Mudd College, 301 Platt Blvd., Claremont, CA 91711}
\author{E.~M.~George} \affiliation{European Southern Observatory, Karl-Schwarzschild-Str. 2, 85748 Garching bei M\"{u}nchen, Germany} \affiliation{Department of Physics, University of California, Berkeley, CA, USA 94720}
\author{A.~Gilbert} \affiliation{Department of Physics, McGill University, 3600 Rue University, Montreal, Quebec H3A 2T8, Canada}
\author{M.~D.~Gladders} \affiliation{Department of Astronomy and Astrophysics, University of Chicago, 5640 South Ellis Avenue, Chicago, IL, USA 60637} \affiliation{Kavli Institute for Cosmological Physics, University of Chicago, 5640 South Ellis Avenue, Chicago, IL, USA 60637}
\author{S.~Guns} \affiliation{Department of Physics, University of California, Berkeley, CA, USA 94720}
\author{N.~Gupta} \affiliation{School of Physics, University of Melbourne, Parkville, VIC 3010, Australia}
\author{N.~W.~Halverson} \affiliation{Department of Astrophysical and Planetary Sciences, University of Colorado, Boulder, CO, USA 80309} \affiliation{Department of Physics, University of Colorado, Boulder, CO, USA 80309}
\author{N.~Harrington} \affiliation{Department of Physics, University of California, Berkeley, CA, USA 94720}
\author{J.~W.~Henning} \affiliation{High Energy Physics Division, Argonne National Laboratory, 9700 S. Cass Avenue, Argonne, IL, USA 60439} \affiliation{Kavli Institute for Cosmological Physics, University of Chicago, 5640 South Ellis Avenue, Chicago, IL, USA 60637}
\author{G.~C.~Hilton} \affiliation{NIST Quantum Devices Group, 325 Broadway Mailcode 817.03, Boulder, CO, USA 80305}
\author{G.~P.~Holder} \affiliation{Department of Physics, University of Illinois Urbana-Champaign, 1110 W. Green Street, Urbana, IL 61801, USA} \affiliation{Astronomy Department, University of Illinois at Urbana-Champaign, 1002 W. Green Street, Urbana, IL 61801, USA} \affiliation{Canadian Institute for Advanced Research, CIFAR Program in Cosmology and Gravity, Toronto, ON, M5G 1Z8, Canada}
\author{W.~L.~Holzapfel} \affiliation{Department of Physics, University of California, Berkeley, CA, USA 94720}
\author{J.~D.~Hrubes} \affiliation{University of Chicago, 5640 South Ellis Avenue, Chicago, IL, USA 60637}
\author{J.~Hubmayr} \affiliation{NIST Quantum Devices Group, 325 Broadway Mailcode 817.03, Boulder, CO, USA 80305}
\author{K.~D.~Irwin} \affiliation{SLAC National Accelerator Laboratory, 2575 Sand Hill Road, Menlo Park, CA 94025} \affiliation{Dept. of Physics, Stanford University, 382 Via Pueblo Mall, Stanford, CA 94305}
\author{G.~Khullar} \affiliation{Kavli Institute for Cosmological Physics, University of Chicago, 5640 South Ellis Avenue, Chicago, IL, USA 60637} \affiliation{Department of Astronomy and Astrophysics, University of Chicago, 5640 South Ellis Avenue, Chicago, IL, USA 60637}
\author{L.~Knox} \affiliation{Department of Physics, University of California, One Shields Avenue, Davis, CA, USA 95616}
\author{A.~T.~Lee} \affiliation{Department of Physics, University of California, Berkeley, CA, USA 94720} \affiliation{Physics Division, Lawrence Berkeley National Laboratory, Berkeley, CA, USA 94720}
\author{D.~Li} \affiliation{NIST Quantum Devices Group, 325 Broadway Mailcode 817.03, Boulder, CO, USA 80305} \affiliation{SLAC National Accelerator Laboratory, 2575 Sand Hill Road, Menlo Park, CA 94025}
\author{A.~Lowitz} \affiliation{Department of Astronomy and Astrophysics, University of Chicago, 5640 South Ellis Avenue, Chicago, IL, USA 60637}
\author{M.~McDonald} \affiliation{Kavli Institute for Astrophysics and Space Research, Massachusetts Institute of Technology, 77 Massachusetts Avenue, Cambridge, MA~02139, USA}
\author{J.~J.~McMahon} \affiliation{Department of Physics, University of Michigan, 450 Church Street, Ann  Arbor, MI, USA 48109}
\author{S.~S.~Meyer} \affiliation{Kavli Institute for Cosmological Physics, University of Chicago, 5640 South Ellis Avenue, Chicago, IL, USA 60637} \affiliation{Department of Physics, University of Chicago, 5640 South Ellis Avenue, Chicago, IL, USA 60637} \affiliation{Department of Astronomy and Astrophysics, University of Chicago, 5640 South Ellis Avenue, Chicago, IL, USA 60637} \affiliation{Enrico Fermi Institute, University of Chicago, 5640 South Ellis Avenue, Chicago, IL, USA 60637}
\author{L.~M.~Mocanu} \affiliation{Department of Astronomy and Astrophysics, University of Chicago, 5640 South Ellis Avenue, Chicago, IL, USA 60637} \affiliation{Kavli Institute for Cosmological Physics, University of Chicago, 5640 South Ellis Avenue, Chicago, IL, USA 60637} \affiliation{Institute of Theoretical Astrophysics, University of Oslo, P.O.Box 1029 Blindern, N-0315 Oslo, Norway}
\author{J.~Montgomery} \affiliation{Department of Physics, McGill University, 3600 Rue University, Montreal, Quebec H3A 2T8, Canada}
\author{A.~Nadolski} \affiliation{Astronomy Department, University of Illinois at Urbana-Champaign, 1002 W. Green Street, Urbana, IL 61801, USA} \affiliation{Department of Physics, University of Illinois Urbana-Champaign, 1110 W. Green Street, Urbana, IL 61801, USA}
\author{T.~Natoli} \affiliation{Department of Astronomy and Astrophysics, University of Chicago, 5640 South Ellis Avenue, Chicago, IL, USA 60637} \affiliation{Kavli Institute for Cosmological Physics, University of Chicago, 5640 South Ellis Avenue, Chicago, IL, USA 60637} \affiliation{Dunlap Institute for Astronomy \& Astrophysics, University of Toronto, 50 St George St, Toronto, ON, M5S 3H4, Canada}
\author{J.~P.~Nibarger} \affiliation{NIST Quantum Devices Group, 325 Broadway Mailcode 817.03, Boulder, CO, USA 80305}
\author{G.~Noble} \affiliation{Department of Physics, McGill University, 3600 Rue University, Montreal, Quebec H3A 2T8, Canada}
\author{V.~Novosad} \affiliation{Materials Sciences Division, Argonne National Laboratory, 9700 S. Cass Avenue, Argonne, IL, USA 60439}
\author{S.~Padin} \affiliation{Kavli Institute for Cosmological Physics, University of Chicago, 5640 South Ellis Avenue, Chicago, IL, USA 60637} \affiliation{Department of Astronomy and Astrophysics, University of Chicago, 5640 South Ellis Avenue, Chicago, IL, USA 60637} \affiliation{California Institute of Technology, MS 249-17, 1216 E. California Blvd., Pasadena, CA, USA 91125}
\author{S.~Patil} \affiliation{School of Physics, University of Melbourne, Parkville, VIC 3010, Australia}
\author{C.~Pryke} \affiliation{School of Physics and Astronomy, University of Minnesota, 116 Church Street S.E. Minneapolis, MN, USA 55455}
\author{C.~L.~Reichardt} \affiliation{School of Physics, University of Melbourne, Parkville, VIC 3010, Australia}
\author{J.~E.~Ruhl} \affiliation{Physics Department, Center for Education and Research in Cosmology and Astrophysics, Case Western Reserve University, Cleveland, OH, USA 44106}
\author{B.~R.~Saliwanchik} \affiliation{Department of Physics, Yale University, 217 Prospect Street, New Haven, CT, USA 06511}
\author{A.~Saro} \affiliation{Astronomy Unit, Department of Physics, University of Trieste, via Tiepolo 11, I-34131 Trieste, Italy} \affiliation{IFPU - Institute for Fundamental Physics of the Universe, Via Beirut 2, 34014 Trieste, Italy} \affiliation{INAF-Osservatorio Astronomico di Trieste, via G. B. Tiepolo 11, 34143 Trieste, Italy}
\author{J.T.~Sayre} \affiliation{Department of Astrophysical and Planetary Sciences, University of Colorado, Boulder, CO, USA 80309} \affiliation{Department of Physics, University of Colorado, Boulder, CO, USA 80309}
\author{K.~K.~Schaffer} \affiliation{Kavli Institute for Cosmological Physics, University of Chicago, 5640 South Ellis Avenue, Chicago, IL, USA 60637} \affiliation{Enrico Fermi Institute, University of Chicago, 5640 South Ellis Avenue, Chicago, IL, USA 60637} \affiliation{Liberal Arts Department, School of the Art Institute of Chicago, 112 S Michigan Ave, Chicago, IL, USA 60603}
\author[0000-0002-7559-0864]{K.~Sharon} \affiliation{Department of Astronomy, University of Michigan, 1085 S. University Ave, Ann Arbor, MI 48109, USA}
\author{C.~Sievers} \affiliation{University of Chicago, 5640 South Ellis Avenue, Chicago, IL, USA 60637}
\author{G.~Smecher} \affiliation{Department of Physics, McGill University, 3600 Rue University, Montreal, Quebec H3A 2T8, Canada} \affiliation{Three-Speed Logic, Inc., Victoria, B.C., V8S 3Z5, Canada}
\author{A.~A.~Stark} \affiliation{Harvard-Smithsonian Center for Astrophysics, 60 Garden Street, Cambridge, MA, USA 02138}
\author{K.~T.~Story} \affiliation{Kavli Institute for Particle Astrophysics and Cosmology, Stanford University, 452 Lomita Mall, Stanford, CA 94305} \affiliation{Dept. of Physics, Stanford University, 382 Via Pueblo Mall, Stanford, CA 94305}
\author{C.~Tucker} \affiliation{Cardiff University, Cardiff CF10 3XQ, United Kingdom}
\author{K.~Vanderlinde} \affiliation{Dunlap Institute for Astronomy \& Astrophysics, University of Toronto, 50 St George St, Toronto, ON, M5S 3H4, Canada} \affiliation{Department of Astronomy \& Astrophysics, University of Toronto, 50 St George St, Toronto, ON, M5S 3H4, Canada}
\author{T.~Veach} \affiliation{Department of Astronomy, University of Maryland College Park, MD, USA 20742}
\author{J.~D.~Vieira} \affiliation{Astronomy Department, University of Illinois at Urbana-Champaign, 1002 W. Green Street, Urbana, IL 61801, USA} \affiliation{Department of Physics, University of Illinois Urbana-Champaign, 1110 W. Green Street, Urbana, IL 61801, USA}
\author{G.~Wang} \affiliation{High Energy Physics Division, Argonne National Laboratory, 9700 S. Cass Avenue, Argonne, IL, USA 60439}
\author[0000-0002-3157-0407]{N.~Whitehorn} \affiliation{Department of Physics and Astronomy, University of California, Los Angeles, CA, USA 90095}
\author[0000-0001-5411-6920]{W.~L.~K.~Wu} \affiliation{Kavli Institute for Cosmological Physics, University of Chicago, 5640 South Ellis Avenue, Chicago, IL, USA 60637}
\author{V.~Yefremenko} \affiliation{High Energy Physics Division, Argonne National Laboratory, 9700 S. Cass Avenue, Argonne, IL, USA 60439}

\correspondingauthor{N.~Huang}
\email{ndhuang@berkeley.edu}

\begin{abstract}
We present a catalog of galaxy cluster candidates detected in 100 square degrees surveyed with the \sptpol\ receiver on the South Pole Telescope.
The catalog contains \ncandidates\ candidates detected with a signal-to-noise ratio greater than \minsig.
The candidates are selected using the Sunyaev-Zel'dovich effect at 95 and 150~GHz.
Using both space- and ground-based optical and infrared telescopes, we have confirmed \nconfirmed\ candidates as galaxy clusters.
We use these follow-up images and archival images to estimate photometric redshifts for \nwithphotoz\ galaxy clusters and spectroscopic observations to obtain redshifts for \nspecz\  systems.
An additional \nconfirmednoz\ galaxy clusters are confirmed using the overdensity of near-infrared galaxies only, and are presented without redshifts.
We find that \nzgtone\ candidates (\pctgtone\ of the total sample) are at redshift of $z \geq 1.0$, with a maximum confirmed redshift of $z_{\rm{max}} = \maxz \pm 0.10$.
We expect this catalog to contain every galaxy cluster with $\mfive > \mcomplete$ and  $z > 0.25$ in the survey area.
The mass threshold is approximately constant above $z = 0.25$, and the complete catalog has a median mass of approximately $\mfive = \massmedian$.
Compared to previous \spt\ works, the increased depth of the millimeter-wave data (11.2 and 6.5 $\mu$K-arcmin at 95 and 150~GHz, respectively) makes it possible to find more galaxy clusters at high redshift and lower mass.
\end{abstract}


\section{Introduction}

As the most massive collapsed objects in the universe, galaxy clusters provide a unique probe of the growth of structure over cosmological time scales (see \citealt{allen11} for a review).
In particular, their abundance is sensitive to the amplitude and shape of the matter power spectrum, as well as the sum of neutrino masses \citep{wang98, wang05, lesgourgues06}, and the nature of the observed cosmic acceleration \citep[e.g.,][]{weinberg13}.
In order to constrain these cosmological parameters, a galaxy cluster sample requires a large extent in redshift, a well understood selection function, and good mass estimates.

A promising method for detecting galaxy clusters was proposed by \citet{sunyaev72}.
They showed that the hot intracluster medium (ICM), which is made up of diffuse plasma at $10^7$ -- $10^8$ K, would cause a spectral shift of the cosmic microwave background (CMB) in the direction of a galaxy cluster.
As CMB photons pass through the ICM, a small fraction inverse-Compton scatter off of the high-energy electrons, boosting them to higher energies.
This causes a decrement in the CMB intensity below 217 GHz, which makes galaxy clusters easy to distinguish from emissive sources.
This has become known as the Sunyaev-Zel'dovich effect (SZE).\footnote{There is an additional effect, known as the kinematic SZE, which is caused by peculiar motions of free electrons.  In this work, we treat this effect as a (small) noise term.}

Compared to traditional methods of finding galaxy clusters (either using the X-ray emission from the ICM, or overdensities in optical/infrared galaxy catalogs), the SZE is less effective at finding low-mass, low-redshift clusters.
On the other hand, the traditional probes rely on intrinsic emission from the galaxy cluster, which is subject to cosmological dimming.
At infrared wavelengths, the K correction compensates for most of the cosmological dimming \citep[see e.g.][and references therein]{gonzalez19}.
However, infrared surveys are affected by projection effects and have higher mass-observable scatter than SZE-selected clusters.
SZE surveys are able to provide approximately mass-limited catalogs across the full redshift range,
with the maximum redshift set by the increasing rarity of high-mass clusters at high redshift.
At the highest redshifts, more massive clusters have not had time to form.

The magnitude of the SZE is given by 
\begin{align}
  \label{eqn:Tsz}
  \Delta T_{\textsc{sz}} &= \Tcmb \fsz\left(x\right) \int n_e \frac{k_B T_e}{m_e c^2} \sigma_{\textsc{t}} dl \\
  &\equiv \Tcmb \fsz\left(x\right) \ysz
\end{align}
\citep{sunyaev72}, where $\Tcmb$ is the temperature of the CMB,
 $x \equiv h \nu / k_B \Tcmb$,
 $n_e$ is the electron number density,
 $T_e$ is the electron temperature,
 $\sigma_{\textsc{t}}$ is the Thomson cross section,
 $k_B$ is the Boltzmann constant,
 $c$ is the speed of light,
 and the integral is along the line of sight.
$\fsz$ describes the frequency dependence of the SZE:
\begin{equation}
  \label{eqn:fsz}
  \fsz\left(x\right) = \left( x \frac{e^x + 1}{e^x - 1}\right) \left(1 + \delta_{\rm{rc}}\right)
\end{equation}
where $\delta_{\rm{rc}}$ is a relativistic correction which is several percent for massive clusters with $T_e > 5$ keV \citep{nozawa00}.
Below 217 GHz, $\fsz < 0$, which leads to the CMB decrement noted above.
Since the SZE is a spectral effect, it is independent of the distance to the galaxy cluster.
Furthermore, \ysz\ is proportional to the thermal energy integrated along the line of sight.
The total SZE signal of the cluster ($Y_{\textsc{sz}}$), defined as the integral of \ysz\ over the transverse extent of the cluster, is mathematically equivalent to the total thermal pressure of the cluster.
It is expected to be tightly correlated with cluster mass \citep{motl05}.
This makes the SZE an effective tool for building galaxy cluster catalogs for cosmological analyses (see \citealt{carlstrom02} for a review).

The SZE decrement only exceeds 100 $\mu\rm{K}_{\rm{CMB}}$\footnote{Here, and throughout this work, values in units of K$_{\textsc{cmb}}$ refer to the equivalent black-body temperature deviation from 2.73 K required to create the observed signal.} for the most massive and rarest clusters, so it is a small signal.
It was not until 2009 that the first previously unknown galaxy cluster was discovered through the SZE \citep{staniszewski09}.
Nonetheless, the last decade has seen the production of SZE-selected galaxy cluster catalogs with hundreds to thousands of objects by the South Pole Telescope (\spt), Atacama Cosmology Telescope (\act), and \planck\ collaborations (e.g. \citealt{bleem15b, hilton18, planck15-27}).

In this paper, we present a catalog of galaxy clusters found in one of the deepest high-resolution CMB maps currently available.
This pushes the cluster detection threshold to lower mass, and represents the first of several catalogs that will use data of similar or greater depth.
By decreasing the mass threshold, we have also increased the effective redshift limit.
The catalog consists of \ncandidates\ galaxy cluster candidates, of which \nconfirmed\ have been confirmed using optical and near-infrared data.
Of the confirmed clusters, \nfirstrep\ are presented for the first time.
This paper is organized as follows:
in \S\ref{sec:mm-wave}, we describe the mm-wave data and processing;
\S\ref{sec:extract} discusses the cluster search methodology and characterization;
\S\ref{sec:oir} describes the optical and infrared follow up used to confirm cluster candidates and estimate their redshifts;
the catalog is presented and compared with other cluster catalogs in \S\ref{sec:catalog};
finally, we discuss conclusions and upcoming work in \S\ref{sec:conclusion}.
Selected data reported in this work (and any updates to the clusters in this catalog) will be available at \webaddress.

\section{Millimeter-wave Data}
The cluster sample presented in this work was derived from two-band millimeter-wave (mm-wave)
data taken with the \sptpol\ receiver on the South Pole Telescope (SPT, \citealt{carlstrom11}).\label{sec:mm-wave}
In this section, we describe the observations and the data processing used to produce mm-wave maps.
\subsection{Telescope and Observations}
\label{sec:obs}

The SPT is a 10-meter-diameter telescope located within 1 km of the geographic South Pole, 
at the National Science Foundation's Amundsen-Scott South Pole Station, one of the premier
sites on Earth for mm-wave observations. The 10-meter aperture results in diffraction-limited
angular resolution of ${\sim}1$ arcmin at 150~GHz, which is well-matched to the angular size
of high-redshift galaxy clusters. Combined with the redshift-independent surface brightness of
the SZE, this makes the SPT a nearly ideal instrument for discovering high-redshift clusters through
the SZE. 
The first receiver on the SPT was used to conduct the 2500-square-degree SPT-SZ survey.
The resultant cluster catalog contains 42 clusters above $z=1$ with typical masses of $3 \times 10^{14} \msunnoh$, the most extensive sample of massive, high-redshift systems selected via ICM observables in the literature \citep[][hereafter B15]{bleem15b}.  

The \sptpol\ receiver \citep{austermann12}, installed on the telescope in 2012, consists of 
1536 detectors, 1176 configured to observe at 150~GHz, and 360 configured to observe at 
95~GHz. The first \sptpol\ observing season and part of the second season were spent observing
a roughly 100-square-degree field centered at right ascension (R.A) 23$^\mathrm{h}$30$^\mathrm{m}$, 
declination -55$^\circ$ (bounded by 23$^\mathrm{h} < $ R.A.~$ < $ 24$^\mathrm{h}$, 
-60$^\circ < $ decl.~$ < $ -50$^\circ$, the same definition as the \textsc{ra23h30dec$-$55} 
field in B15). 
This is referred to as the \sptpol\ 100d field.
It was actually observed as two separate sub-fields: a ``lead'' field and a ``trail'' field.
This strategy was adopted to mitigate the effects of ground-based signals.
No ground signals were ever detected, so we analyze both fields as one.
Observations occurred between March and November of 2012, and in March and April of 2013.
Of approximately 6600 observations of the field, 6150 observations at 95~GHz and 6040 observations at 150~GHz are used in this work.
Maps made from the weighted sum of all \sptpol\ observations of this field have 
rough noise levels (in CMB temperature fluctuation units) of 6.5 $\mu$K-arcmin at 150~GHz
and 11.2 $\mu$K-arcmin at 95~GHz, a factor of 3-4 lower than the typical noise levels
for the SPT-SZ maps used in B15, but over a much smaller area.\footnote{The \sptpol\ receiver is sensitive to both the total intensity
and the polarization of incoming radiation, but we only use the total intensity information in 
this work.} We thus expect a lower total number of clusters compared to B15 but a higher 
density and a larger fraction of systems at high redshift.

\subsection{Mapmaking}
\label{subsec:mapmaking}
The mapmaking procedure followed in this work is nearly identical to that in B15. We summarize
the procedure briefly here and point readers to B15, \citet{reichardt13}, and \citet{schaffer11} 
for more detail. The data from every individual observation of the field at each observing frequency
is calibrated, filtered, and binned into 0.25 arcmin map pixels using
the Sanson-Flamsteed projection \citep{calabretta02}. 
The filtering includes removing the best fit to a set of Legendre polynomials, sines, and cosines, from the time-ordered data of each detector, 
low-pass filtering the data,
and removing the mean and a spatial gradient across the detector array from each time sample.
The equivalent Fourier-domain filtering from each of these steps is a high-pass in the scan direction
with a cutoff of angular multipole $\ell \sim 400$, a low-pass in the scan direction with a cutoff of $\ell \sim 20,000$, and
an isotropic high-pass with a cutoff of $\ell \sim 200$, respectively. 
The low-pass filter is set well above the resolution of the SPT, and is only used to prevent aliasing of high-frequency time-domain noise.
In the two high-pass filtering steps, compact emissive sources with flux density greater than 6.4 mJy at 150~GHz are masked to avoid 
filtering artifacts (the same sources were masked in B15).
 
The filtered, time-ordered data from each detector are inverse-noise weighted, and the weighted data are binned and averaged into pixels. 
The maps from each individual observation are then combined using the total pixel weights into a single full-depth map at each observing frequency.
The pixel weight is the sum of the weights for every time-ordered datum that contributes to the pixel.

\subsection{Beams and Calibration}
\label{sec:cal}

For \sptpol, we measure the instrument response as a function of angle (i.e., instrument beams) 
on planets, particularly Mars. The main lobe of the beam at both 
frequencies is well approximated by an azimuthally symmetric Gaussian, with full width at half maximum (FWHM)
$\sim 1\farcm6$ and $\sim 1\farcm1$ at 95 and 150~GHz respectively. Uncorrected shifts
in absolute pointing between individual-observation maps leads to a further smearing
of the beam in coadded maps, resulting in final beam FWHM of 
$\sim 1\farcm7$ and $\sim 1\farcm2$ at 95 and 150~GHz.
In the matched filter described in the next section, we use a Gaussian approximation
to the beam at each frequency.

The relative calibration among detectors in the \sptpol\ focal plane, and the absolute 
calibration used in this work, are derived using a combination of detector response to an internal thermal source and 
to the Galactic \textsc{Hii} region RCW38. We use a calibration procedure identical to that described
in \citet{schaffer11}, and we refer the reader to that work for details.
We apply an additional calibration factor to our simulations to account for any discrepancies between the RCW38-based calibration and absolute calibration against \sptsz\ maps (see \S \ref{sec:sims}).

\section{Cluster Extraction and Characterization from mm-wave Data}
\label{sec:extract}
In this section, we summarize the procedure used to extract the cluster signal from maps of the microwave sky.
The method used in this work is very similar to that of previous SPT publications.
For a more detailed treatment, see \citet{williamson11}, \citet{reichardt13}, \citet[][hereafter V10]{vanderlinde10}, and B15.

\subsection{Cluster Extraction}
\label{subsec:extraction}

\begin{figure*}[ht]
\centering
\subfigure[95~GHz map]{
  \label{fig:map95}
  \includegraphics[width = .45 \textwidth]{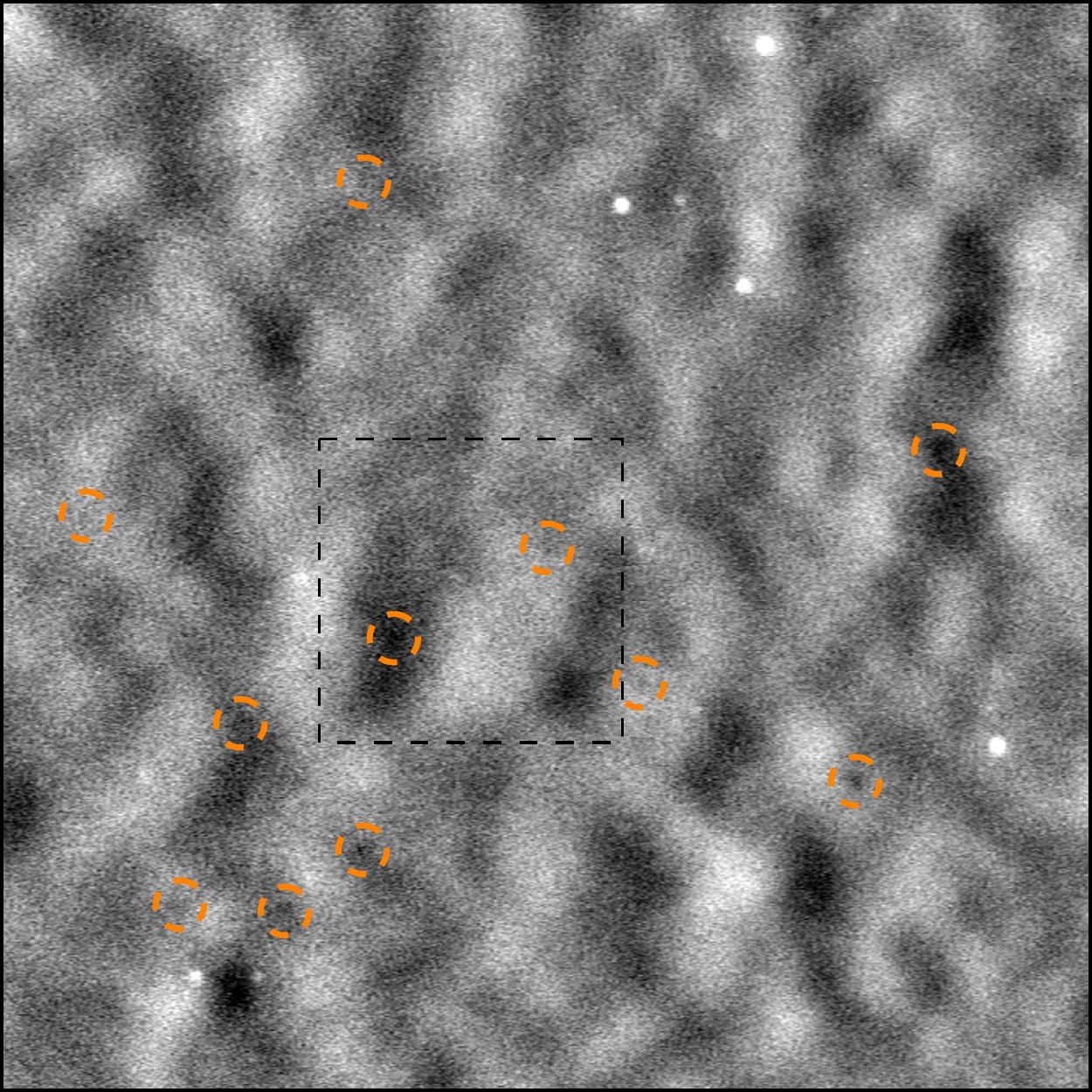}
}
\subfigure[150~GHz map]{
  \label{fig:map150}
  \includegraphics[width = .45 \textwidth]{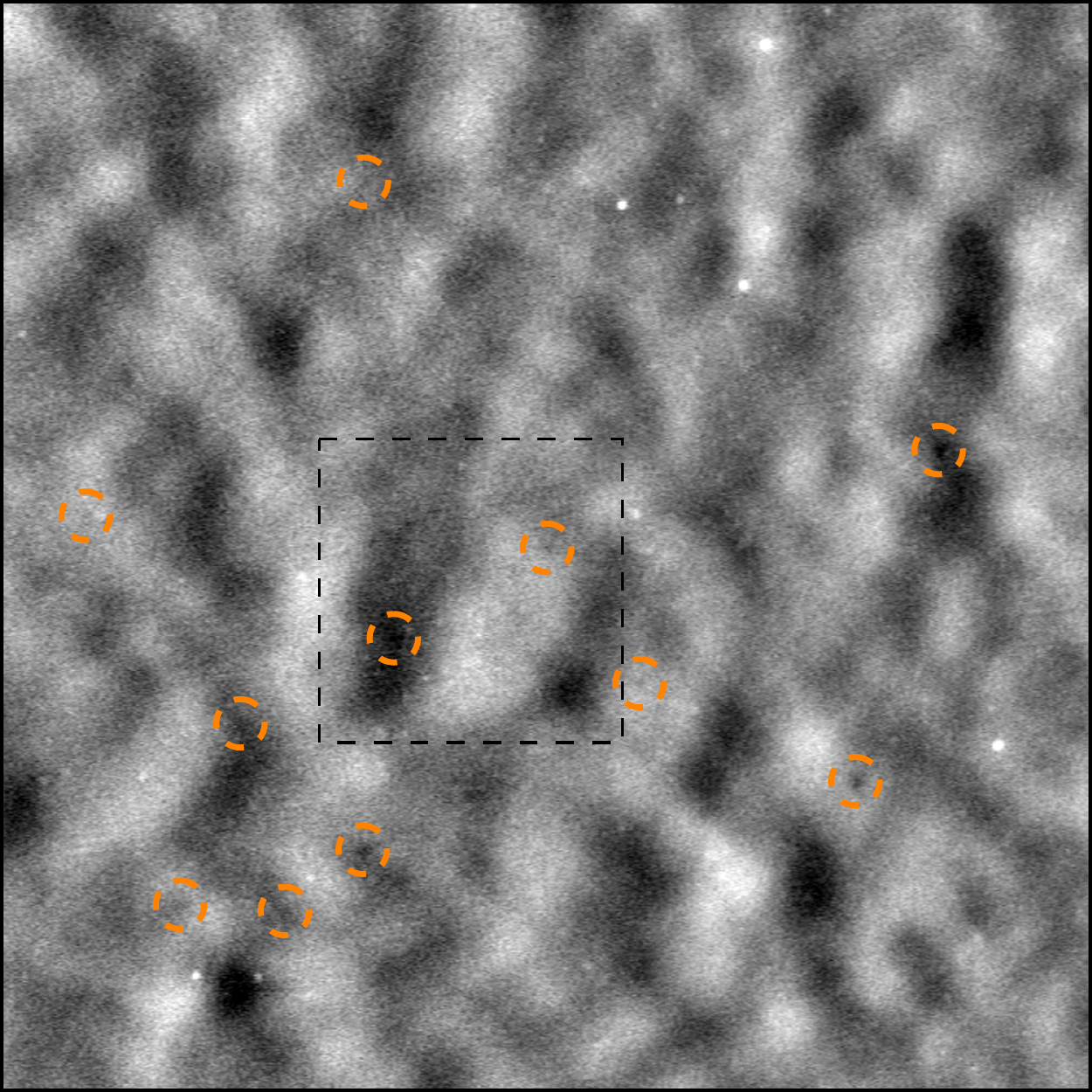}
}
\subfigure[Matched-filtered map]{
  \label{fig:mapfilt}
  \includegraphics[width = .45 \textwidth]{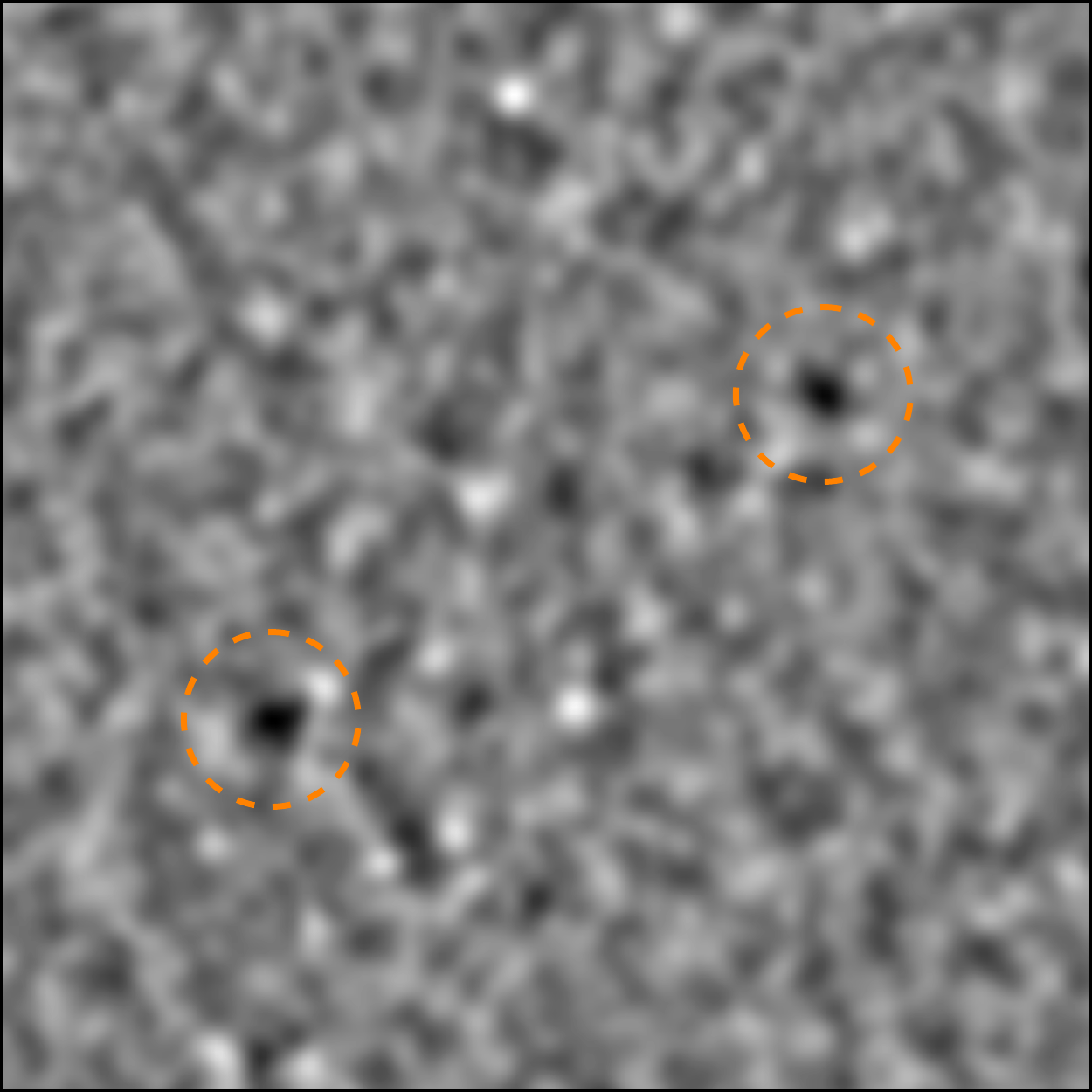}
}
\subfigure[Azimuthally-averaged profile of the matched filter]{
  \label{fig:1dfilt}
  \includegraphics[width = .45 \textwidth]{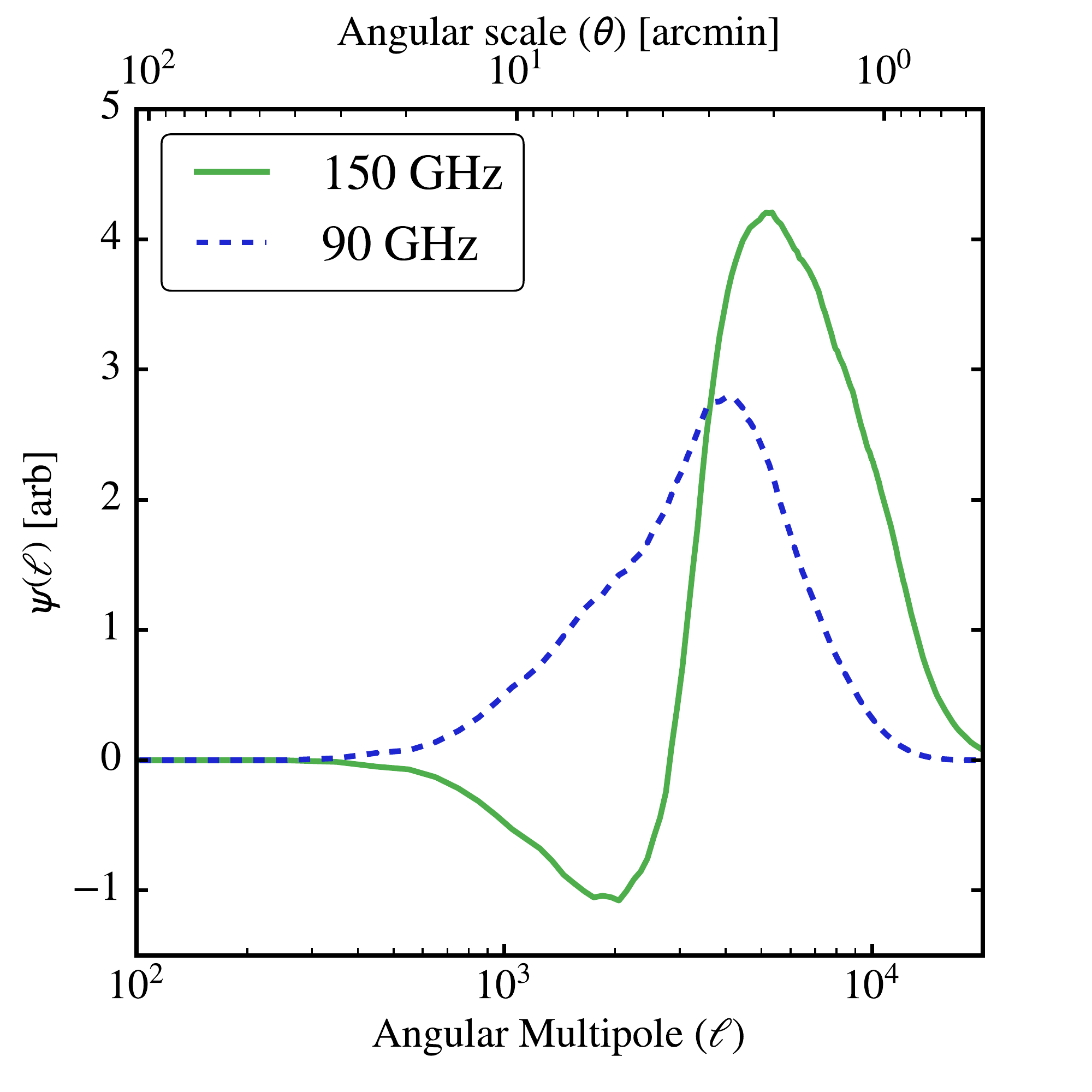}
}
\caption{
  \label{fig:cutouts}
  Panels \subref{fig:map95} and \subref{fig:map150} each show a $3^\circ$-by-$3^\circ$ cutout at 95 and 150~GHz, respectively, with several galaxy cluster detections circled in orange.
  These cutouts are taken from the larger map used in this work, which was produced as described in \S\ref{subsec:mapmaking}.
  Panel \subref{fig:mapfilt} is the region of sky outlined with a black dashed line in panels \subref{fig:map95}, and \subref{fig:map150}, after the matched filter is applied.
  This region is $50\farcm0$ on each side.
  The filtering is described in \S\ref{subsec:extraction}, and this map has been filtered to optimally find objects with $\theta_c = 0\farcm25$.
  It contains 2 detected galaxy clusters, SPT-CL~J2323-5752 at $z = 1.3$, and SPT-CL~J2320-5807 at $z = 0.56$, both of which are detected at a signal-to-noise-ratio greater than 5.0.
  A composite image of SZE contours, optical, and infrared images for SPT-CL~J2323-5752 is shown in Figure \ref{fig:highz}.
  Panel \subref{fig:1dfilt} shows the azimuthally-averaged spatial filter used to produce panel \subref{fig:mapfilt} for each observing band.
Since the SZE has a larger magnitude at 95~GHz, the 95~GHz filter remains positive over the entire range.
The 150~GHz filter is negative at intermediate $\ell$s to subtract CMB fluctuations.
The peak of the 95~GHz filter is lower than the peak of the 150~GHz filter due to the relative noise in the 95 and 150~GHz \sptpol\ maps.
}
\end{figure*}

The \sptpol\ maps contain signals from several classes of sources, each of which has its own spatial and spectral characteristics.
We first describe the spatial behavior of these sources.
At large angular scales, our maps are dominated by the CMB.
On the smallest scales, power comes mostly from point sources, such as dusty galaxies and radio-bright sources.
Galaxy clusters populate the intermediate regime, between the large scale CMB fluctuations and the point sources.

The intensity of these signals also varies by observing frequency.
The amplitude of the CMB and kinematic SZE is preserved across observing frequencies, because the maps are calibrated in CMB temperature units.
Radio-loud active galactic nuclei (AGN) appear with a falling spectrum, while dusty galaxies have a rising spectrum.
Finally, the SZE spectrum is given in equation \ref{eqn:fsz}.
For a given cluster, the magnitude of the SZE is greater at 95~GHz than 150~GHz.
We model the map as follows:
\begin{equation}
  \begin{split}
    T\left(\vec{\theta}, \nu_i\right) = B(\theta, \nu_i) * \left[\Delta T_{\textsc{sz}}\left(\vec{\theta}, \nu_i\right) + N_{\rm{astro}}\left(\vec{\theta}, \nu_i\right)\right] \\
    + N_{\rm{noise}}\left(\vec{\theta}, \nu_i\right)
\end{split}
\end{equation}
where $\vec{\theta}$ is the position on the sky,
$\nu_i$ is the observing band,
$\Delta T_{\textsc{sz}}$ is the SZE, as defined in equation \ref{eqn:Tsz},
$N_{\rm{astro}}$ is all other signals fixed on the sky (such as emissive point sources),
$N_{\rm{noise}}$ is any noise term not fixed on the sky (such as instrumental noise),
$B(\theta, \nu_i)$ represents the effects of the \sptpol\ beam, as well as filtering operations applied during mapmaking,
and $*$ is the convolution operator.

Using the known spatial and spectral forms of the astrophysical noise terms described above, and the measured non-astrophysical noise (see \S\ref{sec:sims}), we construct a simultaneous spatial-spectral filter to optimally extract the cluster signal.
The process is similar to that described in \citet{melin06}.
In the Fourier domain, the filter takes the form
\begin{equation}
  \psi(\mathbf{l}, \nu_i) = \sigma_{\psi}^{2} \sum_j \mathbf{N}_{ij}^{-1}(\mathbf{l}) \fsz(\nu_j) S_{\rm{filt}}(\mathbf{l}, \nu_j)
\end{equation}
where $\sigma_{\psi}^2$ is the predicted variance of the filtered map, given by
\begin{equation}
  \sigma_{\psi}^{-2} = \int d^2l \sum_{i,j}\fsz(\nu_i)S_{\rm{filt}}(\mathbf{l}, \nu_i) \mathbf{N}_{ij}^{-1}(\mathbf{l}) \fsz(\nu_j) S_{\rm{filt}}(\mathbf{l}, \nu_j)
\end{equation}
The band-band, pixel-pixel covariance matrix is represented by $\mathbf{N}$,
and $S_{\rm{filt}}$ is the Fourier transform of the search template convolved with $B(\theta, \nu_i)$.
We use an isothermal projected $\beta$-model \citep{cavaliere76} with $\beta$ fixed at 1 for the SZ surface brightness template:
\begin{equation}
  S = \Delta T_0 \left( 1 + \theta^2 / \theta_c^2\right)^{-\frac{3}{2} \beta + \frac{1}{2}}.
\end{equation}
The normalization ($\Delta T_0$) is a free parameter, and we search over a range of core radii ($\theta_c$). 
V10 explored the use of more sophisticated models, but found no significant improvement.

The covariance matrix $\mathbf{N}$ is constructed using model power spectra for the astrophysical terms, and measured noise properties for the remaining terms.
The astrophysical portion is made up of the lensed CMB, point sources, and kinematic and thermal SZE from unresolved sources.
The CMB power spectrum is calculated using the best-fit WMAP7 + SPT \lcdm\ parameters \citep{komatsu11, keisler11}.
The thermal SZE background is taken to be flat in $l(l+1)$ space, with the level taken from \citet{lueker10}. 
The kinematic SZE spectrum is taken from \citet{shirokoff11}.
While newer results are available for these spectra (e.g., \citealt{george15,planck18-6}), the matched filter is insensitive to the details of the spectra at this level.
The combination of instrumental and atmospheric noise is measured from the data (see \S \ref{sec:sims}).

Before applying the matched filter, we mask point sources above 6.4 mJy (at 150~GHz) with a radius of 4 arcminutes from the source center.
This reduces spurious detections caused by ringing around the brightest sources.
We also apply an additional 8 arcminute veto around these sources, in which we reject any candidate objects.
A small number of spurious detections are still included, which we remove via visual inspection.\footnote{Our visual inspection is statistically consistent with masking point sources above 3 mJy in simulations.}
After masking, the total search area is 94.1 deg$^2$.

We use 12 different filters, with $\theta_c$ evenly spaced between $0\farcm25$ and $3\farcm0$.
After applying the filter in the Fourier domain, we transform back to map space to locate cluster candidates.
Candidates are selected by their signal-to-noise ratio.
Our observation strategy causes small variations in noise in both R.A. and declination.
The overlap region between the lead and trail fields is somewhat deeper than the non-overlapping regions, and instrumental and atmospheric noise are larger closer to the equator.
Our simulations (see \S \ref{sec:sims}) naturally include both noise variations.
We account for the declination-dependence of the noise by splitting the field into $90'$ strips of constant declination and ignore the R.A.-dependence when calculating the signal-to-noise ratio of cluster candidates.
Both variations are small (typicaly $\leq 5\%$), so this will not bias our mass estimation (see \S \ref{sec:mass}).

In each strip, we fit a Gaussian to the distribution of all unmasked pixel values within 5 $\sigma$ of the mean.
We use the standard deviation of the fitted Gaussian as our measure of noise.
The pixel distribution is highly Gaussian within 4 $\sigma$ of the mean, and the fitted standard deviation varies only a small amount when fitting all pixels within 2, 3, 4 or 5 $\sigma$ of the mean.
Therefore, we expect that using the Gaussian approximation of our noise distribution to be sufficiently accurate.
This was changed from previous SPT analyses, which used the root mean square (RMS) of the strip.
For each strip, we divide the filtered map by the noise, resulting in a signal-to-noise map.
The updated method of measuring the noise typically results in a 6-10\% increase in the signal-to-noise ratio at the locations of clusters.
Galaxy cluster candidates are found using a {\tt SExtractor}-like algorithm \citep{bertin96} on each filtered map.
For each candidate, we maximize the signal-to-noise ratio ($\xi$) over R.A., declination, and filter scale.
The matched filter, and the results of applying it to a small section of the maps used in this work are shown in Figure \ref{fig:cutouts}.

\subsection{Simulations}
\label{sec:sims}
In order to characterize our detection algorithm, we use simulated maps of the sky.
Each map has several components:

\begin{description}
\item [CMB] The Code for Anisotropies in the Microwave Background (CAMB; \citealt{lewis00}) is used to generate CMB spectra from the WMAP7+SPT best-fit \lcdm\ parameters \citep{komatsu11, keisler11}.
These are used to generate Gaussian random fields for the CMB realizations.
\item [Radio Sources] We generate point sources at random locations, with fluxes at 150~GHz based on the model from \citet{dezotti05}. 
We assume 100\% spatial correlation between 95 and 150~GHz, and a spectral index $\alpha_{\rm{radio}} = -0.9$ \citep{george15} with Gaussian scatter of 0.47  \citep{mocanu13}.\footnote{We define the spectral index such that the specific intensity of a species $i$ is $I_{\nu,i} \propto \nu^{\alpha_i}$.}
Sources have fluxes between 1.0 and 6.4 mJy.
The upper cutoff was chosen to match the masking threshold used on the real data.
\item [Dusty Sources] Dusty sources are modeled as two Gaussian random fields: one for the Poisson contribution, and another for the clustered term.
We again assume 100\% spatial correlation between bands.
We adopt the model from \citet{george15} and use the best-fit values for spatial and spectral behavior from that work:
the Poisson term is assumed to be of the form $C_\ell = \rm{const.}$, and is normalized such that $D_\ell = \ell \left(\ell + 1\right) / \left(2 \pi\right) C_\ell = 9.16 \, \uk$ at $\ell = 3000$ in the 150~GHz band;
the clustered term is normalized such that $D_{3000} = 3.46 \, \uk$ at 150~GHz, and $D_\ell \propto \ell^{0.8}$.
\citet{george15} finds the best fit spectral indices to be $\alpha_{\rm{clustered}} = 3.27$ and $\alpha_{\rm{poisson}} = 4.27$.
\item [Thermal SZE] N-body simulations are used to project SZE halos onto simulated skies.
Halo locations and masses are taken from the Outer Rim N-body simulation \citep{heitmann19}.
We use the method described in \citet{flender16} to map halos to a location on the sky.
The pressure profile described in \citet{battaglia12} is added at the location of each halo to create a map of the SZE.
The simulated maps contain contributions from halos with $\mfive > 6.25 \times 10^{12} \msun$, spanning a redshift range from $z = 0.01$ to $z = 3.0$.\footnote{\mfive\ is the mass within $r_{500c}$ of the cluster center, where $r_{500c}$ is the radius at which the average density of the galaxy cluster is 500 times the critical density at the cluster's redshift.}
These maps contain a sufficient number of low-mass clusters to reproduce a power spectrum consistent with \citet{george15}.
While the amplitude of the kinematic SZE is similar to that of the thermal SZE, we neglect the kinematic SZE because it is much smaller than other noise terms.
\item [Noise] The instrumental and atmospheric noise terms are not simulated.  Instead, we use noise realizations calculated from the data.  Noise realizations are generated by splitting the maps into equal-weight halves, and then differencing them.  We split the data into halves containing different maps for each noise realization, and normalize based on the number of input maps.
\end{description}

We do not account for any spatial correlation between simulated components.
In particular, this leads to an underestimate of the number of SZE decrements that are partially filled by radio sources.
\citet{bleem19} performs an analysis of the radio intensity at the locations of SZE-selected galaxy clusters, and finds that this is not a large effect.
However, this work uses deeper mm-wave data, so the \citet{bleem19} result should be considered a lower limit to the radio contamination.
The effects of lensing dusty sources and CMB on cluster catalogs was studied in \citet{hezaveh13b}.
They conclude that lensing has little to no effect on the source density in SZE-selected cluster surveys.
Finally, we have not included the kinematic SZE at all.  However, we have tested its impact using the simulations from \citet{flender16}, and find that it adds variance, but no bias. 
We find that the kinematic SZE is responsible for $\sim5\%$ of the intrinsic scatter in the scaling relation between signal-to-noise and mass (see \S \ref{sec:mass}).

These simulations are used to characterize both the expected number of false detections, and the normalization when calculating cluster mass.
In order to properly account for the \sptpol\ beams and calibration, we first filter a map containing all the simulated components with the \sptpol\ filter transfer function and the beam.
For this step, we use the measured beam instead of the Gaussian approximation that is used when constructing the matched filter.
Then, we apply an overall calibration factor to the map for each band.
This calibration factor is calculated via comparisons to calibrated SPT-SZ maps of the same field (see \citealt{keisler15}).
Finally, the instrumental and atmospheric noise is added.

\subsection{Sample Purity}
\label{sec:fdr}
\begin{figure}
  \includegraphics[width = \columnwidth]{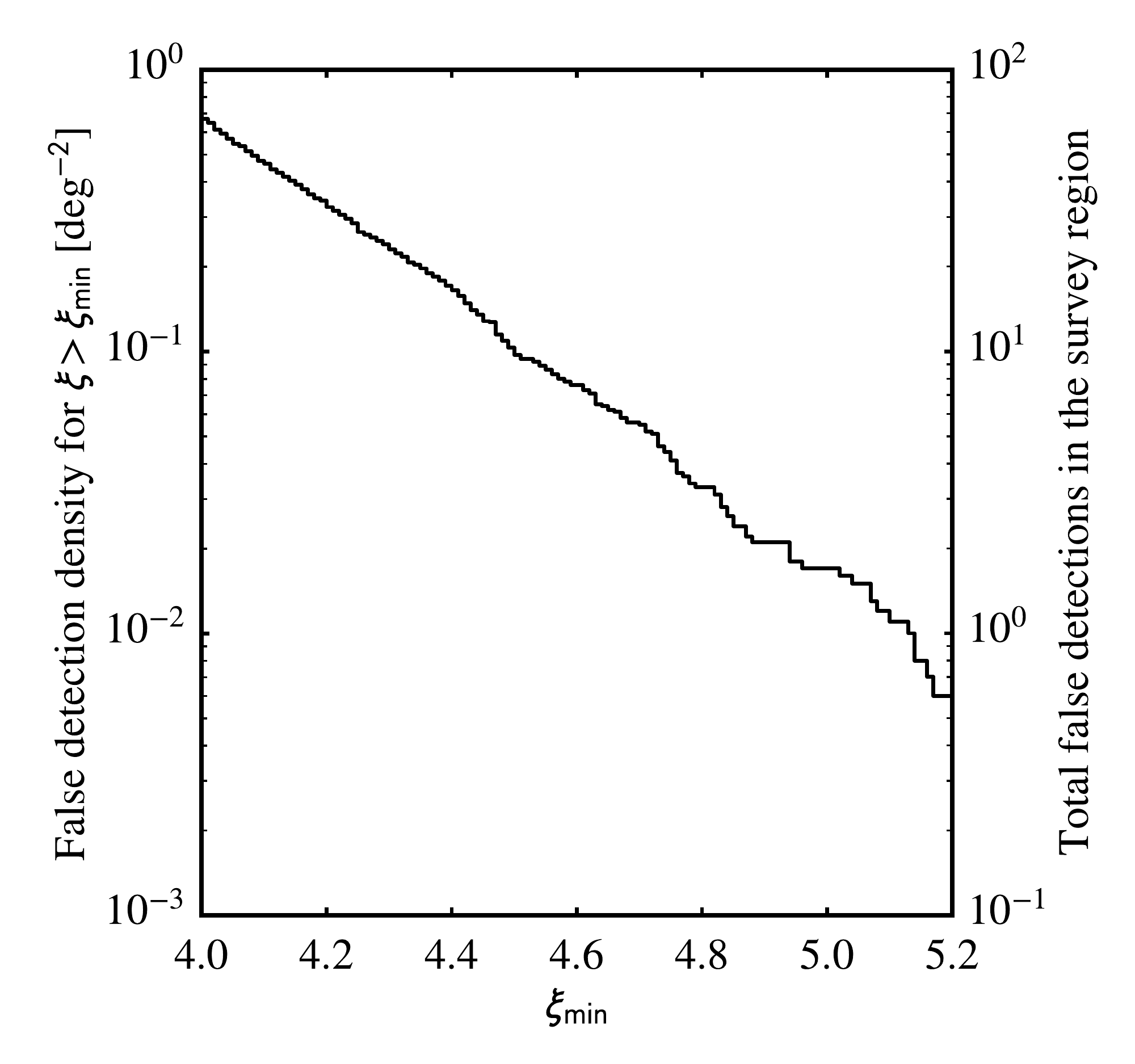}
  \caption{
    \label{fig:fdr}
      Simulated false detection rate, as calculated from approximately 1000 deg$^2$ of simulated sky (10 simulations of the sky area used in this work).
      This shows the density of false detections above a given significance ($\xi_{\rm{min}}$) on the left side, and the expected number of false detections for the entire 100 deg$^2$ survey area on the right.
      We expect $8 \pm 2$ false detections at our significance cutoff, $\xi_{\rm{min}} = \minsig$.
  }

\end{figure}
We expect some fraction of our cluster candidates to be false detections caused by astrophysical and instrumental noise.
In order to estimate the rate of false detections, we create a set of simulated sky maps containing all noise terms, but no SZE signal.
While, in principle, this means we are missing a noise term from sub-threshold galaxy clusters, the power in the SZE is much smaller than other noise terms on the spatial scales that the matched filter is sensitive to.
Furthermore, at current noise levels, chance associations of below-threshold clusters that overlap to form an apparently higher significance cluster are very rare.
We search for clusters in these maps, using the procedure described above (including by-eye cleaning of spurious detections associated with bright point sources).
The remaining detections are assumed to be representative of the false detections we find in the real data.
At our cutoff of $\xi > \minsig$ we expect \fdrpercent\ purity (i.e. \nfdet\ false detections).
See \S \ref{sec:catalog} for more details on the cutoff.
The false detection rate is shown as a function of cutoff significance in Figure \ref{fig:fdr}.

\subsection{Galaxy Cluster Mass}   
\label{sec:mass}
We estimate masses for each cluster based on its detection significance ($\xi$) in SPT maps.
Our method for estimating masses is described in detail in \citet{reichardt13}, \citet{benson13} and \citet{bocquet19}.
In this section, we briefly describe the method.

For each galaxy cluster, we calculate the posterior probability for mass
\begin{equation}
P \left(M | \xi\right) = \left. \frac{dN}{dM dz}\right|_z P \left(\xi | M, z\right)
\end{equation}
where $\xi$ is the measured signal-to-noise ratio of the cluster, $\frac{dN}{dM dz}$ is the assumed mass function \citep{Tinker08}, and $P(\xi|M)$ is the $\xi$-mass scaling relation.
As discussed in V10, $\xi$ itself is a biased mass estimator, so we introduce the unbiased SPT detection significance $\zeta$:
\begin{equation}
\zeta \equiv \sqrt{\left<\xi\right>^2 - 3}.
\end{equation}
For $\left<\xi\right> > 2$, this removes bias caused by the maximization over three degrees of freedom (R.A., declination, and filter scale) performed when producing the catalog.
$\left<\xi\right>$ is the mean significance with which a cluster would be detected in an ensemble of SZE surveys.
For each cluster, we assume that $P(\xi)$ is a Gaussian distribution with unit width, centered on the measured signal-to-noise ratio, and marginalize over the scatter.
Finally, we parametrize the relationship between the unbiased detection significance and cluster mass as
\begin{equation}
\label{eqn:zeta-m}
\zeta = \Asz \left( \frac{\mfive}{3 \times 10^{14} \msun} \right) ^{B_{\textsc{sz}}} 
\left( \frac{E(z)}{E(0.6)} \right) ^{C_{\textsc{SZ}}}
\end{equation}
with normalization $\Asz$, slope $B_{\textsc{sz}}$, redshift evolution $C_{\textsc{sz}}$.
We assume log-normal scatter $\sigma_{\ln\zeta}$ on the $\zeta-M$ scaling relation.

As defined, \Asz\ is dependent on the noise levels in the maps.
In order to use the same scaling relation for a variety of fields, we redefine $\Asz \rightarrow \gamma_{\rm{field}} \Asz$.
To find $\gamma_{\rm{field}}$, we run a modified version of the cluster extraction algorithm.
We filter 10 simulated maps, using all the components (noise and SZE) described above.
We also filter the maps containing only the SZE (signal-only maps).
Instead of running the source extraction on the signal-to-noise ratio maps, we divide the signal-only maps by the noise (calculated as described in \S\ref{subsec:extraction}).
This gives us a direct measure of $\left<\xi\right>$ instead of $\xi$, by removing the scatter associated with the noise terms, and reduces the number of simulations required.
To construct a catalog, we maximize the signal-to-noise over the filter scale at the central pixel of each halo.
We fit the scaling relation using all halos with $\mfive > 2 \times 10^{14} \msun$ and $z > 0.25$, to match the \sptpol\ cluster sample.
Note that the size of the cluster on the sky becomes comparable to CMB fluctuations for galaxy clusters at $z \leq 0.25$, and our mass estimates may be systematically low at these redshifts.

In previous SPT publications, we have calculated $\gamma_{\rm{field}}$ from a common set of simulations.
We use a different set of cluster simulations to calculate $\gamma_{\rm{field}}$ in this work than have been used in previous SPT publications. 
To account for this, we recalculate $\gamma_{\rm{field}}$ for each of the 19 sub-fields in the SPT-SZ observing region using our new simulations, and we compare these results to the SPT-SZ values of $\gamma_{\rm{field}}$ published in \citet{dehaan16}. 
The ratios of the SPT-SZ field scalings in \citet{dehaan16} to those calculated using our new simulations agree at the percent level among the 19 fields, and we take the median of this ratio and apply it to the $\gamma_{\rm{field}}$ we calculate for the SPTpol 100d field. 
This results in a final scaling for the SPTpol 100d field of $\gamma_{\rm{field}} = 2.66$.
Because $\gamma_{\rm{field}}$ is defined relative to V10, this implies that for the same cluster detected in this data and in V10, we should find a value of $\xi$ 2.66 times larger here. A quick check using the most significant object in both catalogs, SPT-CL J2337-5942, bears this out: that cluster is detected in this work with $\zeta = 38.9$ and in V10 with $\zeta = 14.8$ 
(a factor of 2.63 smaller).

In order to provide consistent, comparable mass estimates between cluster catalogs produced by the \spt\ collaboration, we adopt the mass estimation method used in B15. 
In this method, the cosmological parameters are held fixed and only the scaling relation parameters are varied (such that the fitting procedure becomes equivalent to abundance-matching to the fixed cosmology). 
The cluster catalog we use for the scaling-relation fit is the updated B15 catalog presented in \citet{bocquet19}.\footnote{The updated catalog provides better redshift measurements on a handful of clusters and redshifts for a few previously unconfirmed clusters.}
The resulting best-fit parameters are $\Asz = 4.07$, $B_{\textsc{sz}} = 1.65$, $C_{\textsc{SZ}} = 0.63$, and $\sigma_{\ln\zeta} = 0.18$.
As a check, we run the same fit using only the cluster sample presented in this work.
As expected, the constraints are not as tight as those provided by the fit using the B15 sample, but they are consistent.

\section{Optical and IR Followup}
\label{sec:oir}

\begin{figure*}[t]
\begin{center}
  \includegraphics[width = 5.5in]{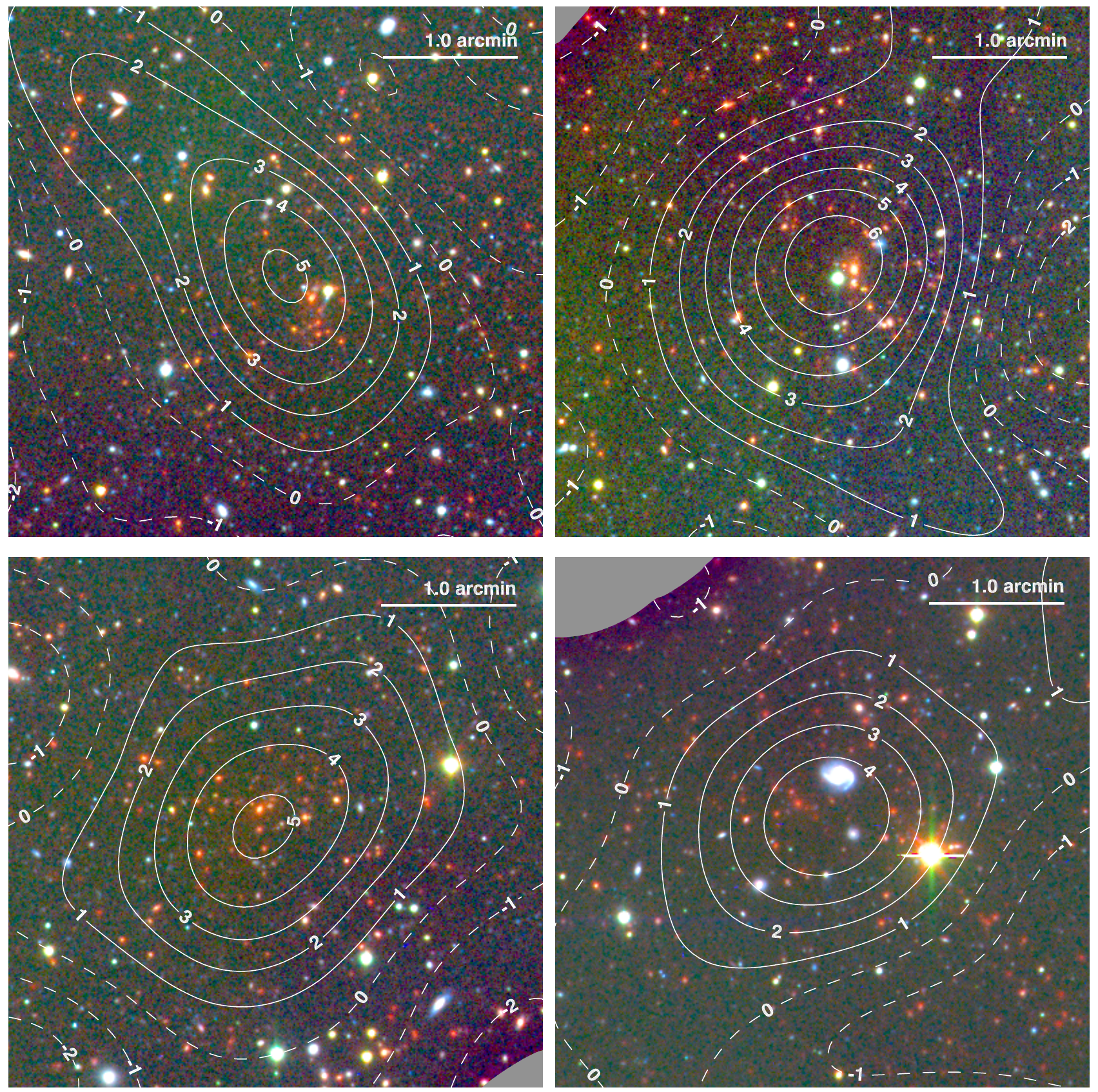}
  \caption{
    \label{fig:highz}
   Four new $z>1$ clusters identified in the \sptpol\ 100d catalog. Clockwise from upper left: SPT-CL~J2259-5301 at $z=1.16 \pm 0.09$ detected at  $\xi=5.1$; SPT-CL~J2336-5252  ($z=1.22 \pm 0.09, \xi=6.7$); SPT-CL~J2323-5752 ($z=1.3 \pm 0.1, \xi =5.1$), and SPT-CL~J2355-5514 ($z=1.4 \pm 0.1, \xi=4.9$).  The composite  \textit{rgb}  images are created using  Magellan/PISCO $g+r$-band data, Magellan/LDSS3-C $z-$band data, and \spitzer \ IRAC 3.6 \um \ data from the SSDF; detection contours from the \sptpol\ $\theta_\textrm{c}$=0\farcm25 filtered maps are overlaid.
}
  \end{center}
\end{figure*}

As in previous SPT cluster publications, we use the existence of overdensities of red-sequence galaxies at the locations of SZ cluster candidates to both confirm candidates as 
clusters as well as to obtain redshift estimates for confirmed systems. Galaxy overdensities are identified in optical and near-infrared imaging observations; such imaging was 
obtained for all 66 cluster candidates at $\xi \ge 5$ and 94\% (\nwithfollowup/\ncandidates) of the candidates at $\xi \ge \minsig$. 
Optical and infrared imaging for four high-redshift clusters is shown in Figure \ref{fig:highz}.

\subsection{Optical Data}

The optical imaging for this work was primarily conducted using the Parallel Imager for Southern Cosmology Observations (PISCO; \citealt{stalder14}), 
a simultaneous 4-band (\textit{griz}) imager with a $9'$ field-of-view mounted on the 6.5-meter Magellan/Clay telescope at Las Campanas Observatory.  
The simultaneous imaging in multiple bands allows PISCO to provide efficient follow-up of faint targeted sources such as galaxy clusters. 

PISCO data were reduced using a custom-built pipeline that incorporates standard image processing corrections (overscan, debiasing, flat-fielding, illumination) 
as well as additional PISCO-specific corrections that account for non-linearities introduced by bright sources. Reduced images are further processed through the 
PHOTPIPE pipeline  \citep{rest05a, garg07,miknaitis07} for astrometric calibration and to prepare for coaddition which is performed with the {\tt SWarp} algorithm \citep{bertin02}. 
Astrometry is tied to bright stars from the Dark Energy Survey public release \citep{des18} and sources from the second Gaia data release \citep{gaia18}. Sources are identified in the coadded 
imaging data utilizing the {\tt SExtractor}  algorithm  \citep{bertin96} (v 2.8.6) in dual-image mode, star-galaxy separation is accomplished using the \textit{SG} statistic introduced in \citet{bleem15a}, 
and photometry is calibrated using the Stellar Locus Regression (SLR; \citealt{high09}). 

Additional optical imaging was drawn from targeted observations with the LDSS3-camera on Magellan/Clay (5 systems) as well as from a 
reprocessing of public data from the Blanco Cosmology Survey (\citealt{bleem15a}; 5 systems). 
These data were processed following the methods described in B15 and \citet{bleem15a}, respectively.  

\subsection{Near-Infrared Data}
\spitzer /IRAC imaging \citep{fazio04} at 3.6~\um\ and 4.5~\um\ is available for the majority (\spitzerpct) of the \sptpol\ 100d sample.  
These data come from both targeted imaging obtained over the course of the SPT-SZ survey (8\%, see B15 for details) as well as from data in the \spitzer-South Pole Telescope Deep Field (SSDF; \citealt{ashby13}).
The 94-square-degree SSDF in particular was designed to overlap with deep observations by the South Pole Telescope in this 100d field, although it does not cover the entire extent of the mm-wave maps used in this work.
Our near-infrared imaging with Spitzer is of sufficient depth to identify galaxy clusters at $z \lesssim 1.5$.

\subsubsection{Near-Infrared Completeness}
\label{subsec:pblank}
We use SSDF galaxy catalogs to estimate the probability that otherwise unconfirmed candidates are false detections.
The method we use was developed in \citet{song12b}; and we only give a brief overview here.
For each cluster candidate, we fit a $\beta$-model with $\beta = 1$ and a background density to the observed number density of galaxies brighter than 18.5 magnitude (Vega) in the 3.6~\um\ channel.
We also perform this fit for 10,000 random locations in the SSDF footprint.
The probability of a candidate being a false detection is then the fraction of the random locations that have a larger $\beta$-model amplitude than the fit at the location of the cluster candidate.
This statistic is reported as $P_{\rm{blank}}$ for all candidates without redshifts and galaxy clusters with $z > 0.7$ in Table \ref{tbl:sptpolcat}.
We consider candidates with $P_{\rm{blank}} < 0.05$ confirmed galaxy clusters.

\subsection{Redshift Estimation}

Cluster candidates are characterized using tools that identify excesses of red-sequence galaxies at cluster locations.
These tools, as well as the models used for the colors of red-sequence galaxies as a function of redshift, are described in detail in B15; here we simply note details specific to the analysis of this \sptpol\ sample.

As this is one of the first analyses to make use of PISCO data for confirming and characterizing SPT cluster candidates, it was necessary to calibrate the cluster red-sequence models to match the PISCO photometric system. 
Further details are given in \citet{bleem19}, which performs a detailed comparison of redshifts derived from PISCO and a sub-sample of the B15 catalog with spectroscopic redshifts.
Similar to other \textit{griz}-band photometry derived from previous targeted observations of SPT clusters, we have transformed the PISCO photometric system to that of the Sloan Digital Sky Survey (SDSS; \citealt{abazajian09}) using instrumental color
terms in the SLR calibration process.\footnote{This transformation works well in the \textit{gri}-bands, but results in a poorer match in the \textit{z}-band, in part because PISCO has a narrower \textit{z}-filter (similar to 
that to be employed on LSST \citealt{ivezic07}). This photometric calibration scheme is sufficient for our work here, where having a well-calibrated color-redshift model is the driving requirement.  However, in the future PISCO data 
will be tied to the LSST photometric system.} 
The PISCO imaging presented here was obtained as part of a broader effort to obtain high-quality data with good seeing on SPT-SZ and \sptpol\ clusters for a number of additional follow-up studies. In the course of this effort, imaging was obtained for a large fraction of galaxy clusters identified in the 2500d SPT-SZ survey with spectroscopic confirmations \citep{ruel14, bayliss16}; 51 of these systems were used to calibrate the red-sequence models for redshift estimation with the PISCO data; from this calibration we determine cluster redshifts to be accurate to $\sigma_z/(1+z) \sim 0.015-0.02$.

Finally, we have also taken advantage of additional spectroscopy of high-redshift clusters obtained following the publication of B15 (e.g., \citealt{bayliss16, khullar19}) to improve our calibration of near-infrared color-redshift models. 

\section{The Cluster Catalog}
\label{sec:catalog}

\begin{figure*}[t]
  \centering
  \includegraphics[width = \textwidth]{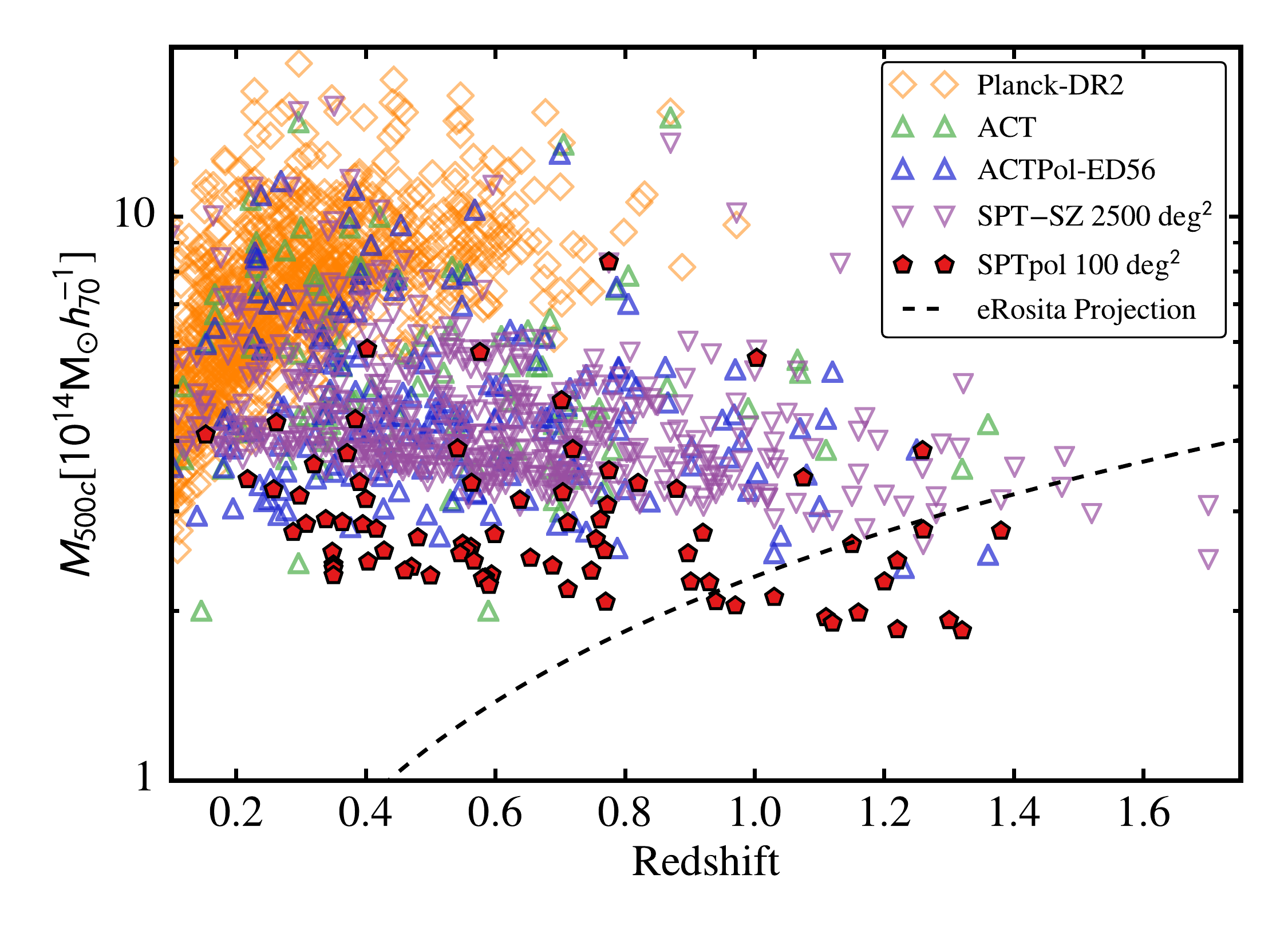}
  \caption{
    \label{fig:m-z}
    The mass and redshift distributions for several recent SZ-selected galaxy cluster catalogs.  
    We plot the estimated mass and redshift for each of the \nwithz\ galaxy clusters with measured redshifts from this catalog, 517 from the \spt\ 2500 deg$^2$ survey (B15), 182 from the \actpol-ED56 survey \citep{hilton18}, 91 from the \act\ survey \citep{hasselfield13}, and 1653 from the second \planck\ galaxy cluster catalog \citep{planck15-27}. 
    The black line is the forecast detection threshold (50 photon counts) for \erosita\ \citep{pillepich12}.}
\end{figure*}

The \sptpol\ 100d cluster catalog contains \ncandidates\ galaxy cluster candidates, \nconfirmed\ of which are optically confirmed and \nfirstrep\ are reported for the first time.
In Table \ref{tbl:sptpolcat}, we present the galaxy cluster candidates found with $\xi > \minsig$.
We originally chose a cutoff of $\xi = 4.5$, so that the marginal false detection rate\footnote{The marginal false detection rate is defined as the fraction of candidates we expect to be false detections in the range $\xi_0 - d \xi < \xi < \xi_0$, where $\xi_0$ is the cutoff significance.} is approximately 50\%.
This choice is largely motivated by the need to follow up our candidates with dedicated optical imaging, and was used to select targets for followup with PISCO.
In the interim, we discovered that it was necessary to change our noise calculation (as noted in \S \ref{subsec:extraction}), which caused $\xi$ for each candidate to increase by approximately 0.1.
This left us with nearly complete followup above $\xi = 4.6$, and a similar marginal false detection rate.

For each candidate, we provide its location, mass, detection significance $(\xi)$, the $\beta$-model core radius $(\theta_c)$ of the filter scale that maximizes detection significance, redshift (if one is available) and $P_{\rm{blank}}$ (see \S \ref{subsec:pblank}, for candidates without redshifts and clusters with $z > 0.7$).
Masses are estimated according to \S\ref{sec:mass}, and redshift estimation is discussed in \S\ref{sec:oir}.
We have \nconfirmednoz\ candidates with $P_{\rm{blank}} \leq 0.05$, but no optical data.
We consider these candidates confirmed clusters, but are unable to provide redshifts or masses at this time.

The median redshift of the catalog is $0.60$, with a maximum redshift of \maxz.
There are \nzgtpeight\ clusters (\pctgtpeight\ of the sample) with $z > 0.8$, \nzgtone\ of which (\pctgtone\ of the sample) are above $z = 1.0$.
The median mass of the catalog is $\mfive = \massmedian$, and spans a range from $\massmin$ to $\massmax$.
The full redshift and mass distribution in comparison to other catalogs is shown in Figure \ref{fig:m-z}.

Based on our simulations, we expect to find $2 \pm 1$ false detections in this catalog (which equates to 98\% purity) above $\xi = 5.0$.
Our optical follow up is consistent with this estimate: it indicates that two of our candidates (SPT-CL~J2349-5140 and SPT-CL~J0002-5214) are likely false detections.
For the full catalog ($\xi > 4.6$) we expect $\nfdet \pm 2$ false detections;  our followup indicates that there are \nfalsedetincat\ false detections.
In addition, \nbrightstar\ candidate cannot be confirmed due to bright stars in the foreground.
Ignoring this unconfirmed candidate, we obtain a catalog purity of \catpurity.
The expected number of false detections is described in detail in \S\ref{sec:fdr} and plotted as a function of minimum significance in Figure \ref{fig:fdr}.
Our optical and near-infrared data on these candidates are of sufficient depth to confirm clusters out to $z \lesssim 1.5$.

\begin{figure}[t]
  \includegraphics[width = \columnwidth]{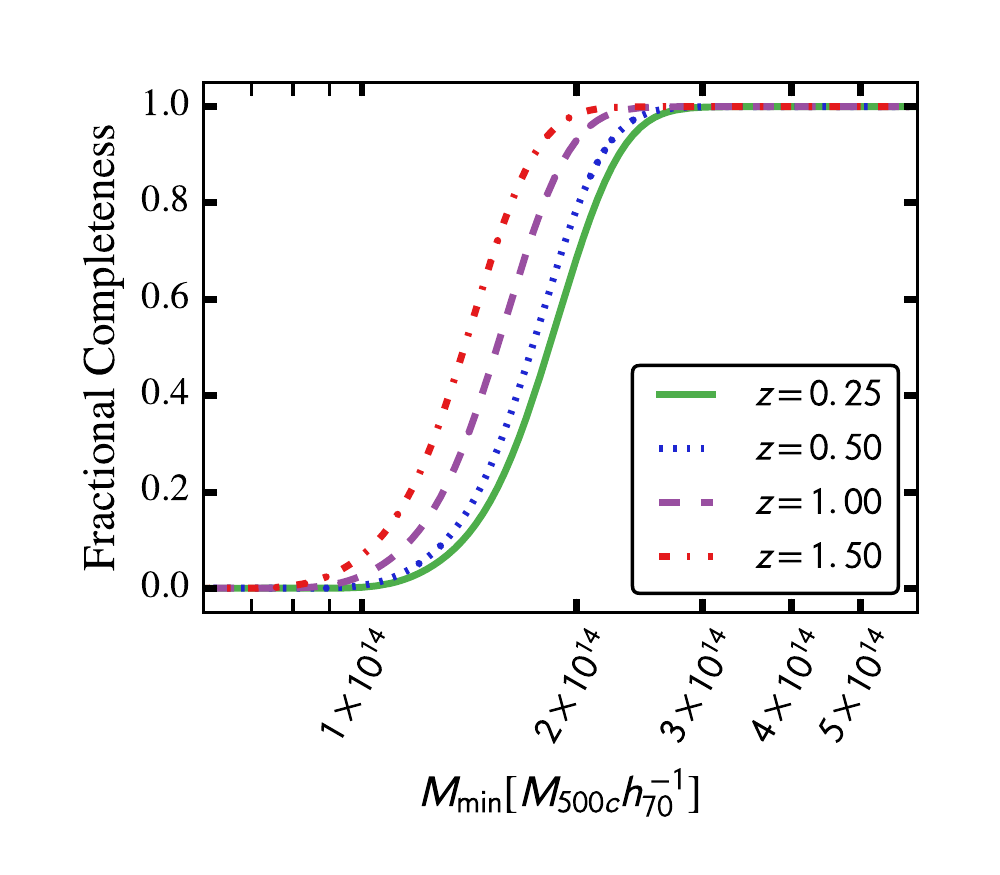}
  \caption{
    \label{fig:selection}
    The estimated completeness as a function of \mfive\ for the \sptpol\ 100d catalog.
    Completeness is estimated at four redshifts: $0.25$, $0.50$, $1.00$, and $1.50$.
    For fixed mass, the completeness at low redshift decreases, because the angular extent of the clusters becomes similar to the fluctuations in the CMB.
    We expect our catalog to be complete for $\mfive > \mcomplete$.
  }
\end{figure}

Figure \ref{fig:selection} shows the estimated completeness for this catalog.
Our selection function is a Heaviside function in significance ($\Theta(\xi - \minsig)$).
We convert this to a completeness in mass and redshift using the $\zeta-M$ scaling relation discussed in \S\ref{sec:mass}.
We find that this catalog is expected to include all galaxy clusters in the survey region more massive than $\mfive = \mcomplete$ and at a redshift greater than $z = 0.25$.

Figure \ref{fig:m-z} shows the mass and redshift distribution for several galaxy cluster surveys.  
The \act, \actpol, \sptsz, and \sptpol\ selection functions are nearly independent of redshift.
There is a slight slope, caused by two effects.
At low redshift, the size of the cluster becomes similar that that of CMB fluctuations and residual atmospheric noise (see e.g., \citealt{schaffer11}).
At moderate to high redshift, the slope is predicted by the self-similar evolution of galaxy clusters.
Self-similarity implies that clusters of the same mass will have higher temperatures at higher redshift, which causes the SZE to be brighter for more distant clusters.
Self-similar evolution further predicts that $C_{\textsc{sz}} = \frac{2}{3}$ \cite{kravtsov06a}.
SZE-selected cluster catalogs generally do not provide strong constraints on self-similarity (\citealt{bocquet19} provides only a $\sim50$\% constraint on $C_{\textsc{sz}}$).
\citet{mcdonald17} used X-ray data to study clusters below $z = 1.9$ and found the bulk of the ICM to be remarkably self-similar.
The \planck\ catalog is limited to lower redshift because the SZ signal from distant clusters is diluted by the relatively large \planck\ beams (7' at 143 GHz).

\subsection{Comparison to Other Cluster Catalogs}
\label{subsec:compare}
In this section, we compare the catalog presented in this work with other cluster catalogs.  
We compare against galaxy clusters in the Simbad\footnote{http://simbad.u-strasbg.fr/simbad/} database, as well as the second \planck\ cluster catalog \citep{planck15-27}, B15, and the \xxl\ 365 catalog \citep{adami18}.
Two objects are considered a match if they fall within $5\farcm0$ of each other below $z = 0.3$.
For higher redshift objects, we apply a threshold of $2\farcm0$ (except for comparisons to the \planck\ catalog, where we use a $4\farcm0$ radius to compensate for the larger \planck\ instrument beam).
The full results can be found in Table \ref{tbl:matches}.
Here, we highlight some recent catalogs.

\begin{figure}
  \centering
  \includegraphics[width = \columnwidth]{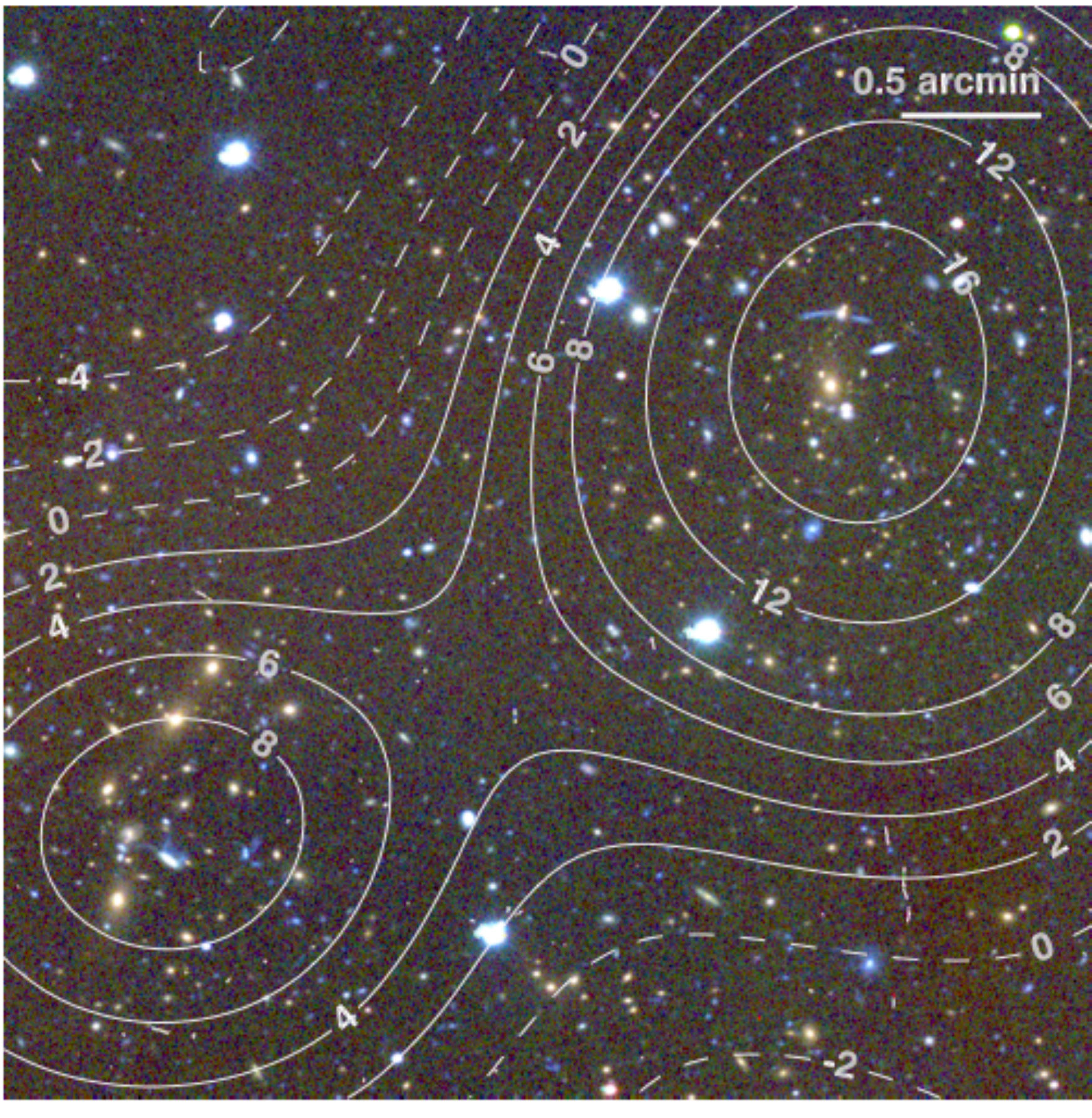}
  \caption{\label{fig:merging}
    A merging cluster, SPT-CL~J2331-5052, identified in the \sptpol\ 100d sample.  This system was detected as two separate clusters in the B15 catalog, but due to the deeper mm-wave data used in this work, it is detected as a single system.  The composite image is created using PISCO \textit{gri} images.  The contours show SZE detection significance from the $\theta_c = 0\farcm5$ match-filtered map.  The position reported in this catalog is in the upper right, near the center of the $\xi = 16$ contour.  In B15, the smaller object (SPT-CL~J2332-5053) is at $z = 0.56 \pm 0.04$; SPT-CL~J2331-5052 is one of the \nspecz\ clusters with spectroscopically measured redshifts at $z = 0.576$. For the more massive object in the upper right of this image, $R_{500} \approx 2\farcm8$.
  }
\end{figure}

The most directly comparable catalog is the \sptsz\ catalog presented in B15.
The full SPT-SZ 2500 deg$^2$ catalog contains \npersqdegsptsz\ confirmed clusters per square degree, while our catalog has nearly four times the density, at \npersqdeg\ clusters per square degree.
The median redshifts of the two catalogs are similar ($z_{\rm{med}} = 0.60$ for this work, and $z_{\rm{med}} =0.55$ in B15), but our catalog contains a much larger fraction at high redshift.
18\% of the clusters with measured redshifts in this work are more distant than $z = 1$, while only 8\% of the clusters in B15 have $z > 1$.

The B15 catalog contains 28 candidates in the \textsc{ra23h30dec$-$55} field, and we find 22 of them in this work.
One cluster (SPT-CL~J2332-5053) is optically confirmed in B15, but not included in this catalog.
It is part of an interacting system with SPT-CL~J2331-5052, which was also noted in previous X-ray observations \citep{andersson11}.
Our source finding algorithm groups all connected pixels above a fixed threshold.
Due to the deeper data used in this work, there is a high significance ``bridge'' that connects the two clusters detected in B15.
By increasing the detection threshold above the minimum value in the bridge in our source finding algorithm, we can force this detection to become separate objects.
We find two clusters, one at (R.A., decl.) of $(352.96122, -50.864841)$ with $\xi = 19.08$, and the other at $(353.02512, -50.892534)$ with $\xi = 9.59$, both with $\theta_c = 0\farcm5$.
Both objects are at the same redshift (the smaller object, reported only in B15 is at $z = 0.56 \pm 0.04$, while SPT-CL~J2331-5052 is at $z = 0.576$).
For the more massive object, $R_{500} \approx 2\farcm8$, which is more than half the projected distance between the two objects.
In Table \ref{tbl:sptpolcat} we report the single cluster SPT-CL~J2331-5052.
This choice is largely arbitrary for this catalog, and we have chosen to be consistent with our stated selection function.
However, future catalogs will find many objects like this, and the choice to report them as single clusters or more than one must be accounted for in any cosmological analyses.
Figure \ref{fig:merging} shows a composite image with combined optical imaging and SZE contours of this system.
The remaining 5 candidates found in B15 are not optically confirmed, which is consistent with the expected false detection rate.
Finally, one candidate (SPT-CL~J2321-5418) is not optically confirmed in this catalog or B15, due to several bright stars in the foreground of the optical imaging.

The sky coverage in this work extends slightly beyond the boundaries of the \textsc{ra23h30dec$-$55} field in B15, and as a result we find an additional 4 clusters which are reported in different fields in the B15 catalog; this brings the total number of matching clusters in both our catalog and B15 to \sptszmatch.
The estimated masses for these two catalogs are consistent with the estimated errors.

The \planck\ galaxy cluster catalog \citep{planck15-27} contains 4 clusters in this field, all of which are included in both this catalog and the \spt\ 2500 deg$^2$ catalog.
The cluster redshifts agree and the scatter in mass between these clusters is consistent with the scatter between the clusters in B15 and the \planck\ cluster catalog.
A more detailed comparison of \planck\ and \sptpol\ masses is included in \citet{bleem19}.

\begin{figure}
  \centering
  \includegraphics[width = \columnwidth]{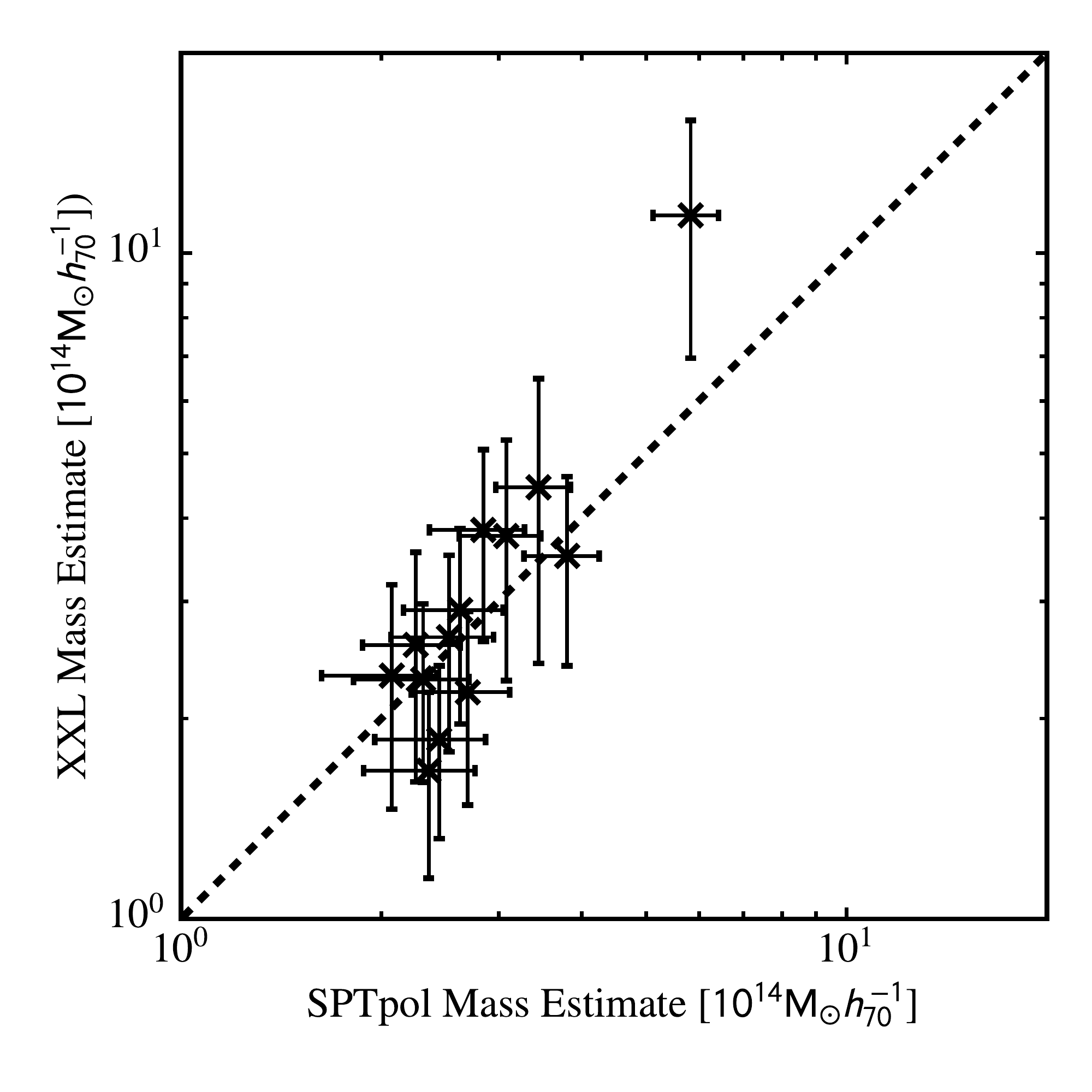}
  \caption{Galaxy cluster masses as measured by \sptpol\ and \xxl.
  Both the \xxl\ and \sptpol\ mass uncertainties are calculated using a scaling relation with an intrinsic scatter in the mass-observable relation (see \S \ref{sec:mass}), which leads to an X-ray or SZ mass uncertainty that is highly correlated between clusters in the sample.
    Only 13 of the 14 clusters found in both surveys are included, because the mass is not provided for one of the \xxl\ clusters.
  }
  \label{fig:xxl}
\end{figure}

The \xxl\ is an X-ray survey performed using \xmm\ \citep{pierre16}, and is one of the deepest X-ray surveys to date.
It covers 50 deg$^2$, 25 of which overlap with the sky area in this work.
In the overlapping area, \xxl\ finds 154 galaxy clusters, the majority of which are at sufficiently low mass or redshift as to be undetectable in this analysis.
14 of these clusters are also found in the \sptpol\ data.
The masses of clusters in the \xxl\ catalog are reported using two different methods.
Since we report masses in \mfive, we compare to the \xxl\ masses calculated using an X-ray temperature scaling relation (since these are also in \mfive).
Of the 14 matching clusters, only 13 have scaling relation masses reported in \citet{adami18}.
We compare the masses of the 13 clusters, and find that they are statistically consistent, with a median ratio of $1.08 \substack{-0.04 \\ + 0.06}$ and RMS scatter of 41\%.
Dividing out the median difference yields a scatter of 38\%.
The scatter is dominated by the highest mass cluster, however the \sptpol\ mass estimate is only $1.2 \sigma$ discrepant from the estimated \xxl\ mass.
The masses of the matching clusters are shown in Figure \ref{fig:xxl}.
We also compare redshifts between the two catalogs.
Most of the redshifts are consistent, but one cluster is significantly discrepant.\footnote{Our optical imaging suggests that there is a very low richness cluster at the reported redshift ($z = 0.81$) of XLSSC 549.  There is also a much larger cluster at higher redshift ($z = 1.2$), which we report as SPT-CL~J2334-5308.  Given that the mass reported for XLSSC 549 is very similar to that of SPT-CL~J2334-5308 ($\mfive = 2.58 \pm 0.97 \times 10^{14} \msun$ and $\mfive = 2.25 \pm 0.38 \times 10^{14}\msun$, respectively), we believe that the reported redshift for XLSSC 549 is incorrect. Assuming that the X-ray temperature remains fixed, the change in redshift would bring the mass of XLSSC 549 closer to the mass measured in this work.}

\section{Conclusions}
\label{sec:conclusion}
This work documents the assembly and properties of the first catalog of galaxy clusters using data from the \sptpol\ receiver.
Using a matched filter on intensity maps from both 95 and 150~GHz bands, we construct a catalog of \ncandidates\ candidates at signal-to-noise ratio \mbox{$\xi>\minsig$}.  
Of these, \nfirstrep\ candidates are reported for the first time, and \nconfirmed\ are confirmed using optical and infrared observations. 
Based on simulations, we expect the sample to be \fdrpercent\ pure (8 false detections), and we find that the number of unconfirmed candidates (\nfalsedetincat) is consistent with this expectation.

We use a program of optical and infrared follow up to both confirm candidates and estimate their redshift.
The median redshift of the catalog is $z \sim 0.6$, with \nzgtpeight\ (\pctgtpeight) clusters at $z > 0.8$, and \nzgtone\ (\pctgtone) at redshift greater than $1.0$.
While this catalog does not contain clusters at the highest redshifts seen in B15 (which contains 6 confirmed clusters at $z \geq 1.4$), the fraction at $z > 1$ is substantially larger.
We estimate galaxy cluster masses based on the detection significance and redshift of each cluster.
Using the fixed set of cosmological parameters and scaling relation from B15, we find masses ranging from $\mfive = \massmin$ to \massmax, with a median mass of \massmedian.

This work will be followed shortly by two more cluster catalogs using data from the \sptpol\ receiver.
The next catalog covers 27 times the area at higher noise levels (the 2700 square degree \sptpol-ECS field, \citealt{bleem19}), while the other covers 5 times the area at similar noise levels (the \sptpol\ 500 square degree field).
This program will continue with the \sptthree\ receiver \citep{benson14}, which began collecting data in early 2017.
The \sptthree\ survey will cover 15 times the area included in this work.
Data from this survey will achieve noise levels of (3.0, 2.2, 8.8) $\mu$K-arcmin at (95, 150, 220) GHz.
These noise levels will allow us to detect over 3000 galaxy clusters with redshifts extending beyond $z = 2.0$, and masses as low as $9 \times 10^{13}\msun$.
On a similar timescale, a new X-ray satellite, \erosita, will survey the entire sky, and produce an extremely large cluster catalog \citep{merloni12}.
Although \erosita\ will detect lower mass clusters, our catalog already includes clusters below the \erosita\ detection threshold at $z>1$.

The sky area used in this work has also been surveyed at several other wavelengths.
In fact, we have used some of those datasets in this work.  
In the infrared, nearly the entire area has been observed by \spitzer /IRAC;
a similar area was observed in the far infrared with the SPIRE instrument on the \textit{Herschel Space Observatory} \citep{pilbratt10};
the Dark Energy Survey \citep{des18} has covered the region in 5 optical and near-infrared bands (\textit{grizY});
the \xxl\ survey has observed 25 deg$^2$ of the \sptpol\ 100d field;
and the Australian Telescope Compact Array (ATCA) was used to survey 86 deg$^2$ of the field at 1.1-3.1 GHz \citep{obrien18}, as well as the entire Southern sky at 20~GHz \citep{ekers07}.
Since these observations are sensitive to different physics (galactic emission in the optical and infrared, ICM temperature in X-rays, and electron pressure in the microwave), they provide semi-independent measures of intrinsic cluster properties.
Combined, this multi-wavelength data set can give us a much clearer picture of the physics at work in the galaxy clusters than any single probe.

\section*{Acknowledgments}

This work was performed in the context of the South Pole Telescope scientific program. 
SPT is supported by the National Science Foundation through grants PLR- 1248097 and 1852617. 
Partial support is also provided by the NSF Physics Frontier Center grant PHY-0114422 to the Kavli Institute of Cosmological Physics at the University of Chicago, the Kavli Foundation and the Gordon and Betty Moore Foundation grant GBMF 947 to  the  University  of  Chicago. 
This  work is  also  supported  by  the  U.S.  Department  of  Energy. 
This material is based upon work supported by the National Science Foundation Graduate Research Fellowship under Grant No. DGE 1752814.
Argonne National Lab, a U.S. Department of Energy Office of Science Laboratory, is operated by UChicago Argonne LLC under contract no. DE-AC02-06CH11357.
We also acknowledge support from the Argonne  Center  for  Nanoscale  Materials. 
B.B. is supported by the Fermi Research  Alliance  LLC  under  contract  no.   De-AC02-07CH11359  with  the  U.S.  Department  of  Energy.
The CU Boulder group acknowledges support from NSF AST-0956135.
The McGill authors acknowledge funding from the Natural Sciences and Engineering Research Council of  Canada, Canadian  Institute  for  Advanced  Research, and the Fonds de Recherche du Qu\'{e}bec Nature et technologies.
The UCLA authors acknowledge support from NSF AST-1716965 and CSSI-1835865.
The Stanford/SLAC group acknowledges support from the U.S. Department of Energy under contract number DE-AC02-76SF00515, and from the National Aeronautics and Space Administration (NASA) under Grant No. NNX15AE12G, issued through the ROSES 2014 Astrophysics Data Analysis Program.
AS is supported by the ERC-StG `ClustersXCosmo' grant agreement 71676, and by the FARE-MIUR grant 'ClustersXEuclid' R165SBKTMA.
The Melbourne group acknowledges support from the Australian Research Council's Discovery Projects scheme (DP150103208).
PISCO observations are supported by NSF AST-1814719.

\facilities{NSF/US Department of Energy 10m South Pole Telescope (SPT):SPTpol, Magellan:Clay (PISCO,LDSS3), Spitzer (IRAC)}

\newpage 
\bibliography{../../BIBTEX/spt}

\clearpage

\appendix
\section{The Cluster Catalog}

In this appendix, we present the full catalog in Table \ref{tbl:sptpolcat}, and clusters with matching objects in external catalogs in Table \ref{tbl:matches}.

\startlongtable
\def\arraystretch{1.2}
\tabletypesize{\footnotesize}

\begin{deluxetable*}{ lcc | cc |c|c|c|c|c}
  \tablecaption{\label{tbl:sptpolcat}
    Galaxy cluster candidates with $\xi \ge \minsig$}
  \tablehead{\\
    \multicolumn{1}{c}{\bf SPT ID} &
    \multicolumn{1}{c}{{\bf R.A.}} &
    \multicolumn{1}{c}{{\bf Decl.}} &
    \multicolumn{2}{c}{\bf Best} &
    \multicolumn{1}{c}{\bf Redshift} &
    \multicolumn{1}{c}{\boldmath \mfive}  &
    \multicolumn{1}{c}{\bf Imaging}  &
    \multicolumn{1}{c}{\boldmath $P_{\rm{blank}}$} &
    \multicolumn{1}{c}{\bf Notes} \\
    \colhead{} &
    \colhead{(J2000)} &
    \colhead{(J2000)}  &
    \colhead{$\xi$} &
    \colhead{$\theta_\textrm{c}$} &
    \colhead{} &
    \colhead{($10^{14} h_{70}^{-1} M_\odot$)} &
    \colhead{}  &
    \colhead{} &
    \colhead{} 
  }

  \startdata
  SPT-CL J0000$-$5748 & $0.2479$ & $-57.8081$ & $14.64$ & $0.25$ & $0.702$ & $4.72 \substack{+0.50 \\ -0.59}$ & 3 & 0.002 & 1 \\
SPT-CL J0000$-$6020 & $0.0323$ & $-60.3405$ & $7.18$ & $0.50$ & $0.762 \pm 0.049$ & $2.90 \substack{+0.39 \\ -0.46}$ & 1 & \nodata & \\
SPT-CL J0001$-$5440 & $0.4132$ & $-54.6695$ & $9.11$ & $0.50$ & $0.820 \pm 0.082$ & $3.37 \substack{+0.40 \\ -0.48}$ & 3 & 0.009 & \\
SPT-CL J0001$-$5614 & $0.4862$ & $-56.2410$ & $5.41$ & $0.25$ & $0.428 \pm 0.036$ & $2.55 \substack{+0.45 \\ -0.46}$ & 3 & \nodata & \\
SPT-CL J0002$-$5017 & $0.6515$ & $-50.2889$ & $5.45$ & $0.25$ & $0.901 \pm 0.033$ & $2.25 \substack{+0.40 \\ -0.41}$ & 1 & \nodata & \\
SPT-CL J0002$-$5214 & $0.5985$ & $-52.2388$ & $5.88$ & $0.25$ & \nodata & \nodata & 3 & 0.445 & 2 \\
SPT-CL J0002$-$5557 & $0.5048$ & $-55.9624$ & $7.22$ & $0.75$ & $1.150 \pm 0.097$ & $2.62 \substack{+0.36 \\ -0.41}$ & 3 & 0.016 & \\
SPT-CL J2259$-$5301 & $344.8284$ & $-53.0308$ & $5.08$ & $0.25$ & $1.160 \pm 0.094$ & $1.98 \substack{+0.34 \\ -0.41}$ & 3 & 0.014 & \\
SPT-CL J2259$-$5349 & $344.7941$ & $-53.8236$ & $7.17$ & $0.25$ & $0.258 \pm 0.023$ & $3.28 \substack{+0.44 \\ -0.52}$ & 3 & \nodata & \\
SPT-CL J2259$-$5431 & $344.9783$ & $-54.5260$ & $7.79$ & $0.75$ & $0.390 \pm 0.043$ & $3.38 \substack{+0.42 \\ -0.51}$ & 3 & \nodata & 3 \\
SPT-CL J2300$-$5331 & $345.1749$ & $-53.5208$ & $10.69$ & $1.00$ & $0.262$ & $4.31 \substack{+0.49 \\ -0.58}$ & 3 & \nodata & \\
SPT-CL J2300$-$5617 & $345.0003$ & $-56.2849$ & $9.53$ & $0.25$ & $0.153$ & $4.11 \substack{+0.47 \\ -0.58}$ & 3 & \nodata & 4 \\
SPT-CL J2301$-$5317 & $345.3371$ & $-53.2843$ & $5.26$ & $1.50$ & $0.348 \pm 0.025$ & $2.54 \substack{+0.44 \\ -0.49}$ & 3 & \nodata & \\
SPT-CL J2301$-$5546 & $345.4486$ & $-55.7759$ & $5.45$ & $0.75$ & $0.748$ & $2.35 \substack{+0.40 \\ -0.44}$ & 3 & 0.005 & \\
SPT-CL J2303$-$5114 & $345.8057$ & $-51.2406$ & $5.72$ & $0.50$ & $0.288 \pm 0.023$ & $2.76 \substack{+0.45 \\ -0.50}$ & 3 & \nodata & \\
SPT-CL J2304$-$5007 & $346.0036$ & $-50.1167$ & $4.79$ & $0.50$ & $0.590 \pm 0.030$ & $2.22 \substack{+0.39 \\ -0.46}$ & 3 & \nodata & \\
SPT-CL J2304$-$5718 & $346.1080$ & $-57.3099$ & $6.25$ & $0.25$ & $0.897 \pm 0.033$ & $2.53 \substack{+0.38 \\ -0.42}$ & 3 & 0.064 & \\
SPT-CL J2305$-$5719 & $346.2706$ & $-57.3261$ & $5.67$ & $0.25$ & $0.654 \pm 0.043$ & $2.48 \substack{+0.43 \\ -0.44}$ & 3 & \nodata & \\
SPT-CL J2306$-$5120 & $346.6121$ & $-51.3465$ & $8.11$ & $0.50$ & $1.260 \pm 0.102$ & $2.78 \substack{+0.36 \\ -0.40}$ & 3 & 0.001 & \\
SPT-CL J2307$-$5440 & $346.7843$ & $-54.6681$ & $5.50$ & $0.75$ & $0.688 \pm 0.044$ & $2.40 \substack{+0.43 \\ -0.43}$ & 3 & \nodata & \\
SPT-CL J2309$-$5710 & $347.2520$ & $-57.1777$ & $6.17$ & $0.25$ & $0.364 \pm 0.035$ & $2.87 \substack{+0.44 \\ -0.49}$ & 3 & \nodata & \\
SPT-CL J2310$-$5239 & $347.7022$ & $-52.6609$ & $4.91$ & $0.50$ & * & \nodata & 2 & 0.047 & \\
SPT-CL J2310$-$5919 & $347.5696$ & $-59.3203$ & $6.12$ & $0.25$ & $0.768 \pm 0.049$ & $2.56 \substack{+0.40 \\ -0.44}$ & 3 & 0.020 & \\
SPT-CL J2311$-$4944 & $347.8888$ & $-49.7397$ & $4.95$ & $1.50$ & $0.584 \pm 0.041$ & $2.29 \substack{+0.41 \\ -0.47}$ & 1 & \nodata & \\
SPT-CL J2311$-$5522 & $347.8989$ & $-55.3724$ & $7.47$ & $0.75$ & $0.217 \pm 0.022$ & $3.42 \substack{+0.45 \\ -0.52}$ & 3 & \nodata & \\
SPT-CL J2311$-$5820 & $347.9955$ & $-58.3363$ & $5.47$ & $0.25$ & $0.930 \pm 0.087$ & $2.25 \substack{+0.40 \\ -0.41}$ & 3 & 0.020 & \\
SPT-CL J2312$-$5101 & $348.2421$ & $-51.0292$ & $4.81$ & $1.00$ & $0.350 \pm 0.024$ & $2.38 \substack{+0.42 \\ -0.49}$ & 3 & \nodata & \\
SPT-CL J2314$-$5554 & $348.5355$ & $-55.9016$ & $4.88$ & $0.25$ & $0.712 \pm 0.044$ & $2.18 \substack{+0.38 \\ -0.46}$ & 3 & 0.074 & \\
SPT-CL J2316$-$5027 & $349.1854$ & $-50.4550$ & $4.73$ & $0.50$ & $1.120 \pm 0.092$ & $1.90 \substack{+0.33 \\ -0.40}$ & 3 & 0.025 & \\
SPT-CL J2316$-$5454 & $349.2113$ & $-54.9020$ & $9.18$ & $1.00$ & $0.371 \pm 0.035$ & $3.80 \substack{+0.44 \\ -0.53}$ & 3 & \nodata & \\
SPT-CL J2317$-$5000 & $349.3259$ & $-50.0018$ & $4.87$ & $0.25$ & $1.110 \pm 0.090$ & $1.95 \substack{+0.34 \\ -0.41}$ & 3 & 0.002 & \\
SPT-CL J2317$-$5357 & $349.3365$ & $-53.9649$ & $6.17$ & $0.25$ & $0.395 \pm 0.035$ & $2.85 \substack{+0.44 \\ -0.48}$ & 3 & \nodata & \\
SPT-CL J2317$-$5852 & $349.4379$ & $-58.8807$ & $5.08$ & $0.50$ & $0.594 \pm 0.042$ & $2.32 \substack{+0.40 \\ -0.47}$ & 3 & \nodata & \\
SPT-CL J2318$-$5059 & $349.5380$ & $-50.9834$ & $4.90$ & $0.75$ & $0.350 \pm 0.024$ & $2.41 \substack{+0.42 \\ -0.49}$ & 3 & \nodata & \\
SPT-CL J2318$-$5617 & $349.7028$ & $-56.2885$ & $5.55$ & $0.25$ & $0.545 \pm 0.039$ & $2.53 \substack{+0.42 \\ -0.46}$ & 3 & \nodata & \\
SPT-CL J2319$-$5842 & $349.8642$ & $-58.7125$ & $6.94$ & $0.50$ & $0.298 \pm 0.023$ & $3.20 \substack{+0.42 \\ -0.50}$ & 3 & \nodata & \\
SPT-CL J2320$-$5233 & $350.1251$ & $-52.5641$ & $6.45$ & $0.25$ & $0.755 \pm 0.046$ & $2.68 \substack{+0.38 \\ -0.44}$ & 3 & 0.001 & \\
SPT-CL J2320$-$5807 & $350.0694$ & $-58.1311$ & $5.80$ & $1.00$ & $0.562 \pm 0.040$ & $2.59 \substack{+0.44 \\ -0.45}$ & 3 & \nodata & \\
SPT-CL J2321$-$5418 & $350.3925$ & $-54.3109$ & $4.68$ & $0.75$ & \nodata & \nodata & 3 & 0.445 & 5 \\
SPT-CL J2323$-$5752 & $350.8787$ & $-57.8831$ & $5.14$ & $0.25$ & $1.300 \pm 0.097$ & $1.92 \substack{+0.35 \\ -0.40}$ & 3 & 0.006 & \\
SPT-CL J2325$-$5116 & $351.3778$ & $-51.2824$ & $5.01$ & $0.25$ & $0.940 \pm 0.082$ & $2.08 \substack{+0.36 \\ -0.43}$ & 3 & 0.002 & \\
SPT-CL J2325$-$5316 & $351.4188$ & $-53.2766$ & $4.60$ & $0.25$ & $0.350 \pm 0.026$ & $2.31 \substack{+0.40 \\ -0.49}$ & 3 & \nodata & \\
SPT-CL J2325$-$5815 & $351.3448$ & $-58.2545$ & $5.72$ & $0.25$ & $0.556 \pm 0.039$ & $2.57 \substack{+0.42 \\ -0.47}$ & 3 & \nodata & \\
SPT-CL J2327$-$5137 & $351.7801$ & $-51.6231$ & $6.22$ & $1.25$ & $0.338 \pm 0.024$ & $2.90 \substack{+0.44 \\ -0.49}$ & 3 & \nodata & \\
SPT-CL J2328$-$5533 & $352.1780$ & $-55.5605$ & $7.88$ & $0.25$ & $0.773 \pm 0.031$ & $3.08 \substack{+0.39 \\ -0.46}$ & 3 & 0.016 & \\
SPT-CL J2328$-$5550 & $352.0383$ & $-55.8466$ & $4.65$ & $0.75$ & $0.770 \pm 0.033$ & $2.07 \substack{+0.35 \\ -0.45}$ & 3 & 0.003 & \\
SPT-CL J2329$-$5831 & $352.4712$ & $-58.5241$ & $10.81$ & $0.50$ & $0.719 \pm 0.045$ & $3.87 \substack{+0.44 \\ -0.52}$ & 3 & 0.001 & \\
SPT-CL J2330$-$5955 & $352.5054$ & $-59.9275$ & $4.90$ & $2.50$ & \nodata & \nodata & 3 & 0.445 & \\
SPT-CL J2331$-$5052 & $352.9767$ & $-50.8720$ & $19.08$ & $0.50$ & $0.576$ & $5.75 \substack{+0.58 \\ -0.70}$ & 3 & \nodata & 1,6 \\
SPT-CL J2331$-$5736 & $352.8991$ & $-57.6143$ & $8.40$ & $0.50$ & $1.380 \pm 0.102$ & $2.77 \substack{+0.34 \\ -0.39}$ & 3 & 0.055 & 7 \\
SPT-CL J2332$-$5220 & $353.1338$ & $-52.3482$ & $4.95$ & $1.00$ & $0.460 \pm 0.026$ & $2.36 \substack{+0.41 \\ -0.47}$ & 3 & \nodata & \\
SPT-CL J2332$-$5358 & $353.1076$ & $-53.9745$ & $18.25$ & $1.50$ & $0.402$ & $5.83 \substack{+0.59 \\ -0.71}$ & 3 & \nodata & 1,8 \\
SPT-CL J2334$-$5308 & $353.5188$ & $-53.1397$ & $6.02$ & $0.25$ & $1.200 \pm 0.094$ & $2.25 \substack{+0.38 \\ -0.38}$ & 3 & 0.001 & \\
SPT-CL J2334$-$5938 & $353.6940$ & $-59.6479$ & $7.09$ & $0.25$ & $0.400 \pm 0.027$ & $3.15 \substack{+0.42 \\ -0.49}$ & 3 & \nodata & \\
SPT-CL J2335$-$5434 & $353.8866$ & $-54.5795$ & $5.26$ & $0.75$ & $1.030 \pm 0.088$ & $2.11 \substack{+0.37 \\ -0.42}$ & 3 & 0.059 & \\
SPT-CL J2336$-$5252 & $354.0877$ & $-52.8725$ & $6.74$ & $0.25$ & $1.220 \pm 0.094$ & $2.45 \substack{+0.35 \\ -0.41}$ & 3 & 0.009 & \\
SPT-CL J2336$-$5352 & $354.0112$ & $-53.8683$ & $5.86$ & $0.75$ & $0.549 \pm 0.039$ & $2.63 \substack{+0.42 \\ -0.47}$ & 3 & \nodata & \\
SPT-CL J2337$-$5912 & $354.3994$ & $-59.2048$ & $6.25$ & $1.00$ & $0.599 \pm 0.041$ & $2.73 \substack{+0.40 \\ -0.46}$ & 3 & \nodata & \\
SPT-CL J2337$-$5942 & $354.3532$ & $-59.7074$ & $38.90$ & $0.25$ & $0.775$ & $8.32 \substack{+0.82 \\ -0.96}$ & 3 & 0.000 & 1 \\
SPT-CL J2339$-$5008 & $354.9618$ & $-50.1427$ & $4.81$ & $0.25$ & * & \nodata & 2 & 0.005 & \\
SPT-CL J2339$-$5550 & $354.8686$ & $-55.8416$ & $5.08$ & $0.50$ & $0.403 \pm 0.036$ & $2.44 \substack{+0.43 \\ -0.49}$ & 3 & \nodata & \\
SPT-CL J2340$-$5958 & $355.0851$ & $-59.9711$ & $4.96$ & $0.25$ & \nodata & \nodata & 3 & 0.758 & \\
SPT-CL J2341$-$5119 & $355.3003$ & $-51.3299$ & $22.05$ & $0.25$ & $1.003$ & $5.61 \substack{+0.58 \\ -0.67}$ & 3 & 0.002 & \\
SPT-CL J2341$-$5138 & $355.4472$ & $-51.6395$ & $5.42$ & $0.75$ & $0.567 \pm 0.040$ & $2.45 \substack{+0.45 \\ -0.45}$ & 3 & \nodata & \\
SPT-CL J2341$-$5640 & $355.4535$ & $-56.6678$ & $5.93$ & $0.25$ & $0.480 \pm 0.038$ & $2.70 \substack{+0.42 \\ -0.48}$ & 3 & \nodata & \\
SPT-CL J2341$-$5724 & $355.3527$ & $-57.4161$ & $13.31$ & $0.50$ & $1.259$ & $3.85 \substack{+0.41 \\ -0.49}$ & 3 & 0.004 & \\
SPT-CL J2342$-$5411 & $355.6940$ & $-54.1869$ & $10.42$ & $0.25$ & $1.075$ & $3.44 \substack{+0.40 \\ -0.47}$ & 3 & 0.022 & 1 \\
SPT-CL J2343$-$5024 & $355.8396$ & $-50.4016$ & $9.03$ & $0.25$ & $0.879 \pm 0.033$ & $3.29 \substack{+0.39 \\ -0.47}$ & 3 & 0.009 & \\
SPT-CL J2344$-$5655 & $356.0958$ & $-56.9292$ & $4.85$ & $0.50$ & $0.500 \pm 0.038$ & $2.31 \substack{+0.41 \\ -0.47}$ & 3 & \nodata & \\
SPT-CL J2349$-$5113 & $357.3841$ & $-51.2259$ & $6.05$ & $0.75$ & $0.416 \pm 0.036$ & $2.79 \substack{+0.45 \\ -0.46}$ & 3 & \nodata & \\
SPT-CL J2349$-$5138 & $357.4703$ & $-51.6417$ & $4.77$ & $0.25$ & \nodata & \nodata & 3 & 0.857 & \\
SPT-CL J2349$-$5140 & $357.2968$ & $-51.6744$ & $5.05$ & $0.25$ & \nodata & \nodata & 3 & 0.887 & \\
SPT-CL J2350$-$5107 & $357.6060$ & $-51.1253$ & $4.63$ & $0.75$ & \nodata & \nodata & 3 & 0.960 & \\
SPT-CL J2350$-$5301 & $357.7262$ & $-53.0218$ & $10.11$ & $0.75$ & $0.541 \pm 0.039$ & $3.88 \substack{+0.45 \\ -0.53}$ & 3 & \nodata & \\
SPT-CL J2351$-$5005 & $357.8298$ & $-50.0936$ & $4.94$ & $0.75$ & $0.580 \pm 0.041$ & $2.28 \substack{+0.40 \\ -0.47}$ & 3 & \nodata & \\
SPT-CL J2351$-$5452 & $357.9044$ & $-54.8825$ & $11.40$ & $0.75$ & $0.384$ & $4.37 \substack{+0.47 \\ -0.58}$ & 3 & \nodata & 1 \\
SPT-CL J2352$-$5251 & $358.2098$ & $-52.8631$ & $5.06$ & $1.00$ & $0.470 \pm 0.037$ & $2.39 \substack{+0.42 \\ -0.48}$ & 3 & \nodata & \\
SPT-CL J2354$-$5106 & $358.5402$ & $-51.1013$ & $6.02$ & $0.50$ & $0.308 \pm 0.024$ & $2.85 \substack{+0.47 \\ -0.48}$ & 3 & \nodata & \\
SPT-CL J2354$-$5632 & $358.7191$ & $-56.5499$ & $8.21$ & $0.75$ & $0.563 \pm 0.040$ & $3.37 \substack{+0.40 \\ -0.49}$ & 3 & \nodata & \\
SPT-CL J2355$-$5055 & $358.9464$ & $-50.9325$ & $8.39$ & $1.00$ & $0.320$ & $3.64 \substack{+0.44 \\ -0.52}$ & 3 & \nodata & \\
SPT-CL J2355$-$5156 & $358.8482$ & $-51.9474$ & $8.24$ & $0.25$ & $0.704 \pm 0.044$ & $3.24 \substack{+0.39 \\ -0.47}$ & 3 & 0.007 & \\
SPT-CL J2355$-$5258 & $358.9372$ & $-52.9833$ & $6.97$ & $0.25$ & $0.712 \pm 0.044$ & $2.87 \substack{+0.40 \\ -0.46}$ & 3 & 0.007 & \\
SPT-CL J2355$-$5514 & $358.8668$ & $-55.2497$ & $4.91$ & $0.25$ & $1.320 \pm 0.098$ & $1.84 \substack{+0.32 \\ -0.40}$ & 3 & 0.005 & \\
SPT-CL J2355$-$5850 & $358.9619$ & $-58.8468$ & $4.94$ & $0.25$ & $0.970 \pm 0.084$ & $2.04 \substack{+0.37 \\ -0.43}$ & 3 & 0.001 & \\
SPT-CL J2355$-$6002 & $358.7942$ & $-60.0428$ & $4.78$ & $0.25$ & $1.220 \pm 0.095$ & $1.85 \substack{+0.32 \\ -0.40}$ & 3 & 0.049 & \\
SPT-CL J2357$-$5421 & $359.2691$ & $-54.3594$ & $7.06$ & $0.25$ & $0.920 \pm 0.081$ & $2.75 \substack{+0.37 \\ -0.43}$ & 3 & 0.014 & \\
SPT-CL J2357$-$5953 & $359.2865$ & $-59.8988$ & $4.67$ & $0.75$ & \nodata & \nodata & 3 & 0.768 & \\
SPT-CL J2358$-$5229 & $359.5318$ & $-52.4840$ & $7.69$ & $0.50$ & $0.638 \pm 0.042$ & $3.15 \substack{+0.41 \\ -0.47}$ & 3 & \nodata & \\
SPT-CL J2359$-$5010 & $359.9302$ & $-50.1708$ & $9.69$ & $0.50$ & $0.775$ & $3.55 \substack{+0.41 \\ -0.50}$ & 3 & 0.003 & 1 \\

  \enddata
  \tablecomments{
The first column is the ID of the candidate.
The second and third column give the position.  
The fourth column gives the signal-to-noise ratio at which the cluster is found, and the fifth column gives the scale of the $\beta$ profile which maximizes the signal-to-noise ratio.
The sixth column lists the redshift, and the seventh column gives the calculated mass for clusters with measured redshift.
The eighth column indicates our available followup for each cluster.
1 means optical imaging only, 2 means near-infrared only, and 3 indicates both optical and infrared imaging.
The ninth column has the statistic $P_{\rm{blank}}$ (the probability that the near-infrared data are consistent with a blank field, see \S \ref{subsec:pblank}) for clusters in the SSDF without measured redshifts, and those with $z > 0.7$.
Redshifts derived from spectroscopic data are reported without uncertainties, but they are typically accurate to $\sim 0.1\%$ \citep[see][]{ruel14, bayliss16}.
The data in this catalog will be hosted at \webaddress.
}
\tablenotetext{1}{Strong-lensing cluster.}
\tablenotetext{2}{There is a group of galaxies at $z = 0.44 \pm 0.04$.}
\tablenotetext{3}{There may be a background cluster at $z \approx 1.1$.}
\tablenotetext{4}{Cluster masses at low redshift ($z < 0.25$) are only approximate. See \S \ref{sec:mass}.}
\tablenotetext{5}{Bright star impedes confirmation.}
\tablenotetext{6}{This cluster is currently undergoing a merger; see \S \ref{subsec:compare}.}
\tablenotetext{7}{There is also a foreground cluster at $z = 0.29 \pm 0.02$.}
\tablenotetext{8}{The mass is biased low by a factor of $\sim 1.5$ owing to contamination from a magnified high-redshift dusty star forming galaxy; see B15 and \citet{andersson11} for details.}
\tablenotetext{*}{Confirmed cluster based on infrared imaging only.  See \S \ref{subsec:pblank}.}
\end{deluxetable*}

\begin{deluxetable*}{l | l | c c c}
  \tablecaption{\label{tbl:matches}
    \sptpol\ clusters matched to other catalogs.
  }    
  \tablehead{
    \colhead{\bf SPT ID} &
    \colhead{\bf Catalogs with Matches} &
    \colhead{\bf \sptpol\ Redshift} &
    \colhead{\bf Lit. Redshift} &
    \colhead{\bf Redshift Ref.}
  }

  \startdata
  SPT-CL J0000$-$5748 & 1 & 0.70 & 0.70 & [1]\\
SPT-CL J0001$-$5440 & 1 & 0.82 & 0.82 & [1]\\
SPT-CL J0002$-$5557 & 1 & 1.15 & 1.15 & [1]\\
SPT-CL J2259$-$5431 & 1,2 & 0.39 & 0.39 & [1]\\
SPT-CL J2300$-$5331 & 1,2,3,4 & 0.26 & 0.26 & [1]\\
SPT-CL J2300$-$5617 & 1,2,3,6,7 & 0.15 & 0.15 & [1]\\
SPT-CL J2301$-$5546 & 1 & 0.75 & 0.75 & [1]\\
SPT-CL J2306$-$5120 & 1 & 1.26 & 1.26 & [1]\\
SPT-CL J2311$-$5522 & 3 & 0.22 & \nodata & \nodata\\
SPT-CL J2311$-$5820 & 1 & 0.93 & 0.93 & [1]\\
SPT-CL J2312$-$5101 & 8 & 0.35 & 0.33 & [8]\\
SPT-CL J2314$-$5554 & 8 & 0.71 & 0.75 & [8]\\
SPT-CL J2316$-$5454 & 1,8,9,10,11,12 & 0.37 & 0.37 & [1]\\
SPT-CL J2317$-$5357 & 12 & 0.40 & 0.38 & [12]\\
SPT-CL J2318$-$5059 & 8 & 0.35 & 0.33 & [8]\\
SPT-CL J2318$-$5617 & 8,9,12 & 0.55 & 0.55 & [12]\\
SPT-CL J2321$-$5418 & 1 & \nodata & \nodata & \nodata\\
SPT-CL J2325$-$5316 & 8,12 & 0.35 & 0.37 & [12]\\
SPT-CL J2327$-$5137 & 1 & 0.34 & 0.34 & [1]\\
SPT-CL J2328$-$5533 & 10,12 & 0.77 & 0.81 & [12]\\
SPT-CL J2328$-$5550 & 10,12 & 0.77 & 0.80 & [12]\\
SPT-CL J2329$-$5831 & 1 & 0.72 & 0.72 & [1]\\
SPT-CL J2331$-$5052 & 1 & 0.58 & 0.58 & [1]\\
SPT-CL J2332$-$5220 & 8,9,12 & 0.46 & 0.46 & [12]\\
SPT-CL J2332$-$5358 & 1,2,7,8,9,12 & 0.40 & 0.40 & [1]\\
SPT-CL J2334$-$5308 & 12 & 1.20 & 0.81 & [12]\\
SPT-CL J2335$-$5434 & 10,12 & 1.03 & 0.67 & [12]\\
SPT-CL J2336$-$5352 & 8,9,12 & 0.55 & 0.52 & [12]\\
SPT-CL J2337$-$5912 & 1 & 0.60 & 0.60 & [1]\\
SPT-CL J2337$-$5942 & 1,2 & 0.78 & 0.77 & [1]\\
SPT-CL J2339$-$5550 & 8,12 & 0.40 & 0.38 & [12]\\
SPT-CL J2341$-$5119 & 1,13 & 1.00 & 1.00 & [1]\\
SPT-CL J2341$-$5640 & 12 & 0.48 & 0.47 & [12]\\
SPT-CL J2341$-$5724 & 1 & 1.26 & 1.26 & [14]\\
SPT-CL J2342$-$5411 & 1,12 & 1.07 & 1.08 & [1]\\
SPT-CL J2350$-$5301 & 1,9 & 0.54 & 0.54 & [1]\\
SPT-CL J2351$-$5452 & 1,8,9 & 0.38 & 0.38 & [1]\\
SPT-CL J2354$-$5632 & 1,7,9 & 0.56 & 0.56 & [1]\\
SPT-CL J2355$-$5055 & 1,7 & 0.32 & 0.32 & [1]\\
SPT-CL J2358$-$5229 & 1 & 0.64 & 0.64 & [1]\\
SPT-CL J2359$-$5010 & 1 & 0.78 & 0.77 & [1]\\

  \enddata

  \tablecomments{Cluster candidates with matches in other published catalogs.
We take two objects to be a match if they are within $2\farcm0$ ($5\farcm0$) if their redshift is higher (lower) than $z = 0.3$, except \planck\ clusters, which are considered to be a match within $4\farcm0$ at $z > 0.3$.
When available, we provide the redshift of the matching cluster.
}
\tablerefs{
[1] \citet{bleem15b};
[2] \citet{voges99};
[3] \citet{abell89};
[4] \citet{skrutskie06};
[5] \citet{jones09};
[6] \citet{braid78};
[7] \citet{planck15-27};
[8] \citet{bleem15a};
[9] \citet{menanteau10};
[10] \citet{suhada12};
[11] \citet{mauch03};
[12] \citet{adami18};
[13] \citet{pascut15};
[14] \citet{khullar19};
}

\end{deluxetable*}

\end{document}